\documentclass[12pt, letter]{emulateapj}
\usepackage{epsf}
\usepackage{graphicx, commath, color, listings}

  \newcommand{\photoz}{photo-$z$}
  \newcommand{\photozs}{photo-$z$s}
  \newcommand{\specz}{spec-$z$}
  \newcommand{\speczs}{spec-$z$s}

  \newcommand{\msun}{M_{\odot}}

  \newcommand{\synmag}{{\sc synmag}}
  \newcommand{\birth}{$b_{1000}$}
  \newcommand{\logm}{\log{M_*/\msun}}

\newcommand{\area}{$139.4$}   
\newcommand{\vol}{$0.3$}   
\newcommand{\cat}{{\sc s82-mgc}}   
\newcommand{\ukwide}{{\sc ukwide}}   

\newcommand{\comppaper}{Paper II}   
\newcommand{\redmapper}{{\it redMaPPer}}  
\newcommand{\redmagic}{{\it redMaGiC}}  
\newcommand{\EAZY}{{EAZY}}  
\newcommand{\BPZ}{{BPZ}}  
\newcommand{\kcorrect}{{KCORRECT}}  
\newcommand{\isedfit}{\texttt{iSEDfit}}

\newcommand{\mangle}{{\sc mangle}}  
\newcommand{\idlutils}{{\sc idlutils}}  
\newcommand{\zbest}{$z_{\rm best}$} 
\newcommand{\zreis}{$z_{\rm Reis}$}
\newcommand{\zeazy}{$z_{\rm EAZY}$}
\newcommand{\zbpz}{$z_{\rm BPZ}$}

\shorttitle{S82-MGC I: Catalog Construction}
\shortauthors{Bundy et al.}

\submitted{Submitted to ApJS}

\begin{document}

\title{The Stripe 82 Massive Galaxy Project I: Catalog Construction}

\author{Kevin Bundy, Alexie Leauthaud, Shun Saito\altaffilmark{1}, Adam Bolton\altaffilmark{2}, Yen-Ting
Lin\altaffilmark{3}, Claudia Maraston, Robert C.~Nichol\altaffilmark{4}, Donald P.~Schneider\altaffilmark{5,6}, Daniel Thomas\altaffilmark{4}, David A.~Wake\altaffilmark
{7,8}}

\altaffiltext{1}{Kavli Institute for the Physics and Mathematics of the Universe (WPI),
   The University of Tokyo Institutes for Advanced Study, The University of Tokyo, Kashiwa, Chiba 277-8583, Japan
}
\altaffiltext{2}{Department of Physics and Astronomy, University of Utah, 115 S 1400 E, Salt Lake City, UT 84112, USA}
\altaffiltext{3}{Institute of Astronomy and Astrophysics, Academia Sinica, Taipei 106, Taiwan}
\altaffiltext{4}{Institute of Cosmology and Gravitation, University of Portsmouth, Portsmouth, UK}
\altaffiltext{5}{Department of Astronomy and Astrophysics, The Pennsylvania State University, University Park, PA 16802, USA}
\altaffiltext{6}{Institute for Gravitation and the Cosmos, The Pennsylvania State University, University Park, PA 16802, USA}
\altaffiltext{7}{Department of Physical Sciences, The Open University, Milton Keynes, MK7 6AA, UK}
\altaffiltext{8}{Department of Astronomy, University of Wisconsin-Madison, 475 N. Charter Street, Madison, WI, 53706, USA}

\begin{abstract}

  The Stripe 82 Massive Galaxy Catalog (\cat{}) is the largest-volume stellar mass-limited sample of galaxies beyond $z
  \approx 0.1$ constructed to date. Spanning \area{} deg$^2$, the \cat{} includes a mass-limited sample of 41,770
  galaxies with $\log{M_*/\msun}
   \gtrsim 11.2$ to $z \approx 0.7$, sampling a volume of \vol{} Gpc$^3$, roughly equivalent to the volume of the
  Sloan Digital Sky Survey-I/II (SDSS-I/II) $z < 0.15$ \textsc{main} sample.  The catalog is built on three pillars of
   survey data: the SDSS Stripe 82 Coadd photometry which reaches $r$-band magnitudes of $\sim$23.5 AB, $YJHK$
   photometry at depths of 20$^{\rm th}$ magnitude (AB) from the UK Infrared Deep Sky Survey Large Area Survey, and
   over 70,000 spectroscopic galaxy redshifts from the SDSS-I/II and the Baryon Oscillation Spectroscopic Survey.
   We describe the catalog construction and verification, the production of 9-band matched aperture photometry, tests
   of existing and newly estimated photometric redshifts required to supplement spectroscopic redshifts for 55\% of the
   $\log{M_*/\msun}
   \gtrsim 11.2$ sample, and geometric masking.  We provide near-IR based stellar mass estimates and compare
  these to previous estimates.  All catalog products are made publicly available.  The \cat{} not only addresses
  previous statistical limitations in high-mass galaxy evolution studies but begins tackling inherent data challenges
  in the coming era of wide-field imaging surveys.

\end{abstract}


\section{Introduction}\label{intro}

A new era of truly panoramic imaging surveys has begun that promises new insights into fundamental questions in
cosmology and galaxy evolution from $z \lesssim 1$ that have been hindered by the relatively small volumes and
resulting statistical uncertainty available to date.  Advancing the legacy of the Sloan Digital Sky Survey
\citep[SDSS,][]{york00}, surveys like DES\footnote{Dark Energy Survey}, HSC\footnote{Hyper-Suprime Cam Survey},
KiDS\footnote{Kilo-Degree Survey}, VIKINGS\footnote{VISTA Kilo-degree Infrared Galaxy Survey}, DECaLS\footnote{Dark
Energy Camera Legacy Survey, \texttt{http://legacysurvey.org/decamls}}, and eventually Euclid, LSST\footnote{Large
Synoptic Sky Telescope}, and WFIRST will provide thousands if not tens of thousands of square degrees of deep multiband imaging data.

These data sets offer exciting opportunities.  Beyond cosmological constraints, from the galaxy evolution perspective,
they will enable precise measurements of evolving number densities that can chart galaxy growth and the rate
of flow between transitioning populations, thus constraining the physical drivers of evolution.  They also present new
challenges, from questions of how to self-consistently process and analyze huge data volumes, to the rising
importance of subtle systematic errors in the estimators we use to derive (photometric) redshifts, total fluxes,
intrinsic colors, and physical properties such as star formation rate and stellar mass, $M_*$.

Ahead of the rising tide of ``Big Data'' in astronomy, we can already make progress on ``large-volume'' questions and
begin tackling some of the challenges above using extant surveys.  The Canada-France-Hawaii Telescope Legacy Wide
Survey (CFHTLS-Wide) is a pioneering example in the $\sim$100 deg$^2$ regime.  The ``Stripe 82'' region, spanning a
2\fd5 wide by 110 degree long stretch of the celestial equator in the Southern Galactic Cap, is another example.  As we
describe below, Stripe 82 is not as deep as CFHTLS-Wide, but offers two features that we exploit in this paper to
construct the Stripe 82 Massive Galaxy Catalog (\cat{}), the largest-volume near-IR selected $M_*$-limited sample of
galaxies beyond $z \approx 0.1$ assembled to date.

First, Stripe 82 was the subject of repeated imaging in SDSS and therefore reaches $ugriz$ depths that are roughly 2
magnitudes deeper (90\% completeness for galaxies at $r \sim 22.5$ AB, \citealt{annis14}) than the single-epoch
imaging.  An ``SDSS Coadd'' of these data was processed and analyzed to produce a publicly available catalog as part of
SDSS Data Release 7
\citep{abazajian09, annis14}.  \citet{jiang14} provide an alternate set of Coadd images, but no photometric catalog.
The added depth in the Coadd is critical for obtaining reliable photometric redshifts (\photozs{}) for massive galaxies
($\log M_*/M_{\odot} > 11$) that can be used to supplement the color-selected spectroscopic samples out to $z \sim
0.7$.  Second, enabling more robust stellar mass estimates and better measures of spectral energy distributions,
near-IR photometry in Stripe 82 is available from the UKIRT Infrared Deep Sky Survey
\citep[UKIDSS,][]{lawrence07}.  Specifically we use the $YJHK$ photometry from the UKIDSS Large Area Survey (LAS)
component which reaches a depth of AB$\sim$20 over roughly 230 deg$^2$ in Stripe 82.

Stripe 82 has also featured prominently in spectroscopic campaigns.  The SDSS-I/II \textsc{main} sample, the
luminous red galaxy (LRG) sample \citep{eisenstein01}, and the Baryon Oscillation Spectroscopic Survey \citep[BOSS][]{dawson13} as well as 2SLAQ, 2dF,
6dF, and WiggleZ \citep{drinkwater10} all cover Stripe 82.  In addition, the stripe has been observed by narrower but deeper surveys such as VVDS
\citep[VIMOS VLT Deep Survey,][]{le-fevre05}, the DEEP2 Galaxy Redshift Survey \citep[][]{davis03,newman13}, and PRIMUS
\citep[PRIsm MUlti-object Survey,][]{coil11}.  This array of spectroscopic redshifts is a critical scientific
asset to the \cat{}. The large number of spectroscopic redshifts (\speczs{}) from SDSS (especially BOSS) provides the
foundation needed to build a complete, $M_*$-limited sample. Roughly 45\% of galaxies in the \cat{} sample with $\logm
> 11.2$ have
\speczs{}.  Deeper \speczs{} provide additional checks and training sets for photometric redshifts (\photozs{}) that
are required to supplement incomplete \specz{} samples in the \cat{}.

A predecessor to the \cat{} was presented in a pioneering effort by \citet{matsuoka10} who downloaded and
reprocessed the UKIDSS-LAS (DR3) $K$-band imaging data and a large fraction of the SDSS Coadd $ugriz$ imaging in a 55
deg$^2$ region of overlap in Stripe 82.  The UKIDSS sources were smoothed to a similar spatial resolution as the SDSS
and photometry performed with Source Extractor
\citep{bertin96}.  \citet{matsuoka10} use their catalog to constrain the growth history of massive galaxies, finding
evidence for a more rapid increase since $z \sim 0.9$ in the number of the most massive galaxies ($\log{M_*/\msun} >
11.5$) compared to a less massive sample ($11.0 < \logm < 11.5$).  The construction of the \cat{} was motivated by
similar goals which will be the subject of a future paper.  In addition to the nearly three times larger area covered
by the \cat{} and the large number of \speczs{} now available, a major difference of our approach compared to \citet
{matsuoka10} is the use of the \synmag{} synthetic aperture photometric matching technique that works at the catalog
level without requiring reprocessing of imaging data \citep[see][]{bundy12}.  With the necessary catalog information in
hand, \synmag{} results can be obtained quickly for data volumes of nearly any size, making 9-band matched photometry over the
full Stripe 82 relatively easy to achieve.  


As the \cat{} was being assembled, new surveys in Stripe 82  started on several facilities: CS82 (the CFHT Stripe 82
Survey, Kneib et al., in preparation), an $i$-band imaging survey to $i_{\rm AB} \approx 23.5$ covering 173 deg$^2$
with a median seeing of 0\farcs6; the VISTA-CFHT Stripe 82 Survey (VICS82, Geach et al., in preparation), which uses
the Visible Infrared Survey Telescope for Astronomy (VISTA) and the CFHT Wide-field InfraRed Camera to image 140
deg$^2$ in the $J$ and $K_s$ bands to 22 AB (5$\sigma$ point-source); and the Spitzer-IRAC Equatorial Survey (SpIES,
PI: Richards), which has now mapped 100 deg$^2$ in the Spitzer 3.6$\mu m$ and 4.5$\mu m$ channels to 5$\sigma$ depths
of 22.8 AB and 22.1 AB, respectively. Adding information from these observations and others to \cat{} will be valuable.
Another valuable dataset is the near-IR photometry of WISE \citep[Wide-field Infrared Survey Explorer][]{wright10}. New
catalogs of WISE+SDSS matched photometry are now available \citep{lang14}. The WISE W1 and W2 channel depths are
roughly comparable to the UKIDSS LAS. Here we focus on the UKIDSS photometry, especially for the additional near-IR
measurements that are useful for constraining galaxy SEDs.

In this paper, the first in a series, we describe the construction of the Stripe 82 Massive Galaxy Catalog, its
underlying datasets, various tests and validation efforts we have undertaken, and value-added products we have derived.
The \cat{} is publicly available from {\texttt {massivegalaxies.com}}.  The \synmag{} tool
necessary for fast, catalog-level matched aperture photometry in the \cat{} was presented in \citet{bundy12} and makes
use of Gaussian Mixtures as described in \citet{hogg13}.  Paper II in the series \citep{leauthaud15}
uses the \cat{} to characterize the $M_*$ completeness of the BOSS spectroscopic samples.  Paper III (Bundy et
al.~2016) analyzes mass-complete samples drawn from the \cat{} to set constraints on evolution in the galaxy stellar
mass function and growth among the most massive galaxies since $z
\sim 0.7$.  The \cat{} is also used in \citet{saito15} in their characterization of the host dark matter
halos of BOSS galaxies and in Chabonnier et al.\ (in preparation) in a study of highly compact massive galaxies.

The paper is structured as follows.  Section \ref{data} summarizes the SDSS Coadd and UKIDSS imaging as well as
the spectroscopic samples that form the basis of the \cat{}.  The application of \synmag{}s to derive matched
photometry is described in Section \ref{photometry}, as is our methodology for building new total flux measurements in
the near-IR given the significant problems in the UKIDSS-LAS estimators.  Star-galaxy separation is discussed in
Section \ref {stargal}, where near-IR colors are used to provide a significant improvement over a classification based
on SDSS shapes alone.  We compile and test photometric redshifts in Section \ref{photoz}.  In Section \ref{geometry} we
analyze the geometry of the \cat{}'s on-sky footprint using the \mangle{} software package and apply rejection masks.
A clean, near-IR limited subsample, \ukwide{}, is presented and characterized in Section \ref{sec:ukwide}.  Section \ref
{properties} describes our derived measurements of galaxy properties, including $M_*$ estimates, which we compare
against previous values produced by the BOSS team.


Throughout this paper, we use the AB magnitude system \citep{oke83} and adopt a standard cosmology with $H_0$=70
$h_{70}$ km s$^{-1}$ Mpc$^{-1}$, $\Omega_M$=0.3 and $\Omega_{\Lambda}$=0.7.  Stellar mass estimates assume a Chabrier
initial mass function \citep[IMF,][]{chabrier03} unless otherwise noted.

\section{Survey Data}\label{data}

The \cat{} is built on three pillars of survey data---the SDSS Stripe 82 Coadd catalog, the UKIDSS LAS, and the BOSS
spectroscopic redshift samples.  We summarize each of these in this section.  With the SDSS Coadd as the starting
point, the \cat{} consists of 15,342,585 objects, including both extended and point sources.  After matching to UKIDSS
photometry (Section \ref{synthetic_mags}), applying cuts in limiting depths (Section \ref{ukidss_depth}), improved star
galaxy separation (Section \ref{stargal}), and geometric masking (Section \ref{geometry}), we derive an $M_*$-limited
sub-sample of \cat{} that we refer to as \ukwide {} (Section \ref{sec:ukwide}).  The \ukwide{} sample consists of
517,714 galaxies, with 41,770 above the nominal completeness limit of$\log {M_*/\msun} > 11.2$.  The publicly available
(go to \texttt{massivegalaxies.com})
input and derived data products associated with the \cat{} are summarized in the Appendix.

\subsection{SDSS Stripe 82 Coadd catalog}

The ``SDSS Coadd'' refers to the stacked imaging and photometric catalog in Stripe 82 (-50\arcdeg{} $<$ $\alpha_{\rm
J2000}$ $<$ +60\arcdeg{}) first presented in \citet{abazajian09} and further described in \citet{annis14}.  The Coadd
was made by processing $\sim$90 repeated visits to the stripe during the SDSS-I/II imaging campaign, including
observations obtained during the SDSS Supernova Survey
\citep{frieman08}.  The source detection,
flagging, profile fitting, and photometry was performed on data taken before 2005 by
\citet{annis14} by applying the {\sc Photo} pipeline \citep{lupton02}.  The SDSS camera is described in
\citet{gunn98}, the telescope system in \citet{gunn06}, and the filter system in \citet{fukugita96}.  The 50\% point-source completeness limit of the SDSS Coadd is $r \sim 24.4$
(AB) and the 90\% galaxy completeness limit is $r \sim 22.5$ (AB).  We used the SDSS Catalog Archive Server\footnote{\texttt{http://skyserver.sdss.org}} and the \texttt{PhotoPrimary} view to query
all sources in the Coadd photometric catalog, \texttt{Stripe82}, with \texttt{TYPE} values of 3 or 6.  The resulting
catalog has 15,342,585 sources and serves as the basis of the \cat{}.  Please see \citet{annis14} and the SDSS website
for further details on executing the query.

A number of checks on the Coadd photometry were performed by \citet{annis14}.  In building the \cat{}, we also compared
the colors and the magnitudes as measured in the SDSS Coadd to those from the Canada-France-Hawaii Telescope Lensing
Survey (CFHTLenS) photometry catalog \citep{hildebrandt12} for sources overlapping the CFHT W4 field.  We found
excellent agreement and no evidence for biased photometry in the Coadd.  Further support comes from the quality of
template-based
\photozs{} measured using the Coadd photometry (Section \ref{sub:template_photozs}).  Compared against \speczs{} from
VVDS, template-based redshift estimators (see Section \ref{photoz}) applied to the Coadd delivers \photozs{} of
comparable quality ($\sigma_z/(1+z) \approx 0.05$ with an outlier fraction of $\sim$10\%) to those reported for
CFHTLenS \citep{heymans12}.  This is despite the fact that the PSF-homogenized CFHTLenS aperture photometry reaches
more than a magnitude deeper than the Coadd profile-fit photometry.

\subsection{UKIDSS Large Area Survey}

The Large Area Survey (LAS) component of UKIDSS is described by \citet{lawrence07} and utilizes the Wide Field Camera
(WFCAM) on the 3.8 m United Kingdom Infrared Telescope (UKIRT).  The LAS is taking advantage of the large WFCAM field
of view (0.21 deg$^2$) to image 4000 deg$^2$ in the $YJHK$ filter set.  As of UKIDSS Data Release 10, the LAS was
roughly two-thirds complete in terms of area covered. The depth goals in AB magnitudes are $Y = 20.9$, $J=20.4$,
$H=20.0$, and $K=20.1$.  We discuss UKIDSS depths further in Section \ref{sec:completeness}.



To construct the \cat{}, we obtained public UKIDSS-LAS Data Release 8 (DR8) catalogs from the WFCAM Science
Archive\footnote {http://surveys.roe.ac.uk/wsa} (WSA).  The LAS covers the majority of Stripe 82 and was
substantially complete in this region in DR8, containing 5,072,574 near-IR selected sources, although some holes remained.  

When aperture magnitudes are required, we use UKIDSS magnitudes computed in the 4$^{\rm th}$ aperture option
(\texttt{AperMag4}) with a radius of 2\farcs8, unless otherwise noted.  Aperture magnitudes in UKIDSS catalogs have
been adjusted to ``total magnitudes'' under the assumption that every object is a point-source. To handle extended
sources, a complex query (see Appendix) is required to obtain the ``correction'' (e.g., \texttt{AperCor4}) so that it
can be removed. We also apply Galactic extinction corrections by subtracting the UKIDSS tabulated \texttt{AX} values,
(where \texttt{X} can be $YJHK$).

The WFCAM focal plane consists of four separate detectors arranged in a checkerboard pattern.  UKIDSS obtains large
survey coverage by tiling this pattern across the sky.  In any given position, multiple short exposures are coadded
into a data unit referred to as a \emph{multiframe}.

\subsubsection{UKIDSS imaging depth and seeing} \label{ukidss_depth}

The UKIDSS imaging depth varies with location in different ways for different bands.  We use the LAS catalog
information to estimate the depth of the imaging in which each source was detected and measured as follows.  First we
identify the set of sources in a given band belonging to a specified \emph{multiframe}.  We then compare the {\tt
AperMag} magnitudes against their magnitude errors and interpolate to find the magnitude at which the average error is
0.1 mag.  We consider this a rough estimate of the 10$\sigma$ depth of that particular \emph{multiframe} and associate
this depth limit with all sources belonging to that \emph{multiframe}.  Some \emph{multiframes} do not have enough
detections to provide a solid depth estimate; this situation is often an indication of a problem in the UKIDSS data.
Sources from such \emph{multiframes} are assigned a magnitude depth of 0.0.  The depth information is used to define a
flux-limited survey area in Section \ref{sec:ukwide}.

We obtain the seeing measurement (FWHM) in each UKIDSS band from the \texttt{MultiframeDetector} tables on the WSA. 
See the Appendix for the relevant query.

\subsection{SDSS Spectroscopic Redshifts}

Our primary source of spectroscopic redshifts is from BOSS, an SDSS-III program \citep[][]{eisenstein11} that obtained
spectroscopic redshifts for 1.5 million galaxies over 10,000 deg$^2$ using the BOSS spectrographs \citep{smee12} on the
2.5 m Sloan Foundation Telescope at Apache Point Observatory \citep[][]{gunn06}. An overview of the BOSS survey is
presented in \citet{dawson13}. BOSS galaxies were selected from SDSS Data Release 8 imaging \citep[][]{aihara11} using
a series of color and magnitude cuts. A concise review of the selection criteria for the two BOSS samples is presented
in \comppaper{}.  The LOWZ sample targeted luminous red galaxies (LRGs) at $0.15<z<0.43$, while the CMASS sample aimed
to collect massive galaxies in a broader color range at $0.43<z<0.7$. Redshifts were measured using the processing
pipeline described in \citet{bolton12}.  We also include redshifts of SDSS-I/II ``Legacy'' objects, primarily
LRGs.

All SDSS redshifts in the \cat{} were obtained by cross-matching to the SpecObj-dr10 catalog\footnote{Available at
\texttt{http://www.sdss3.org/dr10/tutorials/lss\_galaxy.  No additional spectra were made available in the relevant
fields in DR12.}} using a 0\farcs8 tolerance. Matching is done only to specObj entries that have SPECPRIMARY=1 and
ZWARNING\_NOQSO=0 (for BOSS \texttt{programNames}) and ZWARNING=0 (for non-BOSS
\texttt{programNames}).  Both stars and galaxies are matched, producing a total of 149,439 spectroscopic redshifts included in
the \cat{}.

\section{Photometry}\label{photometry}

\subsection{Matched $ugrizYJHK$ \synmag{} Photometry}\label{synthetic_mags}

Photometry matched to a consistent point-spread function (PSF) across the SDSS+UKIDSS filterbands is a key component of the \cat{} that enables SED fitting and
near-IR based $M_*$ estimates.  We describe our use of catalog-level matching \synmag{}s in this section.

As described in \citet{bundy12}, consistent photometry across multiple filter bands with
varying PSFs (and especially from multiple telescopes) is most often obtained by convolving the images available in each
band to the poorest PSF in the ensemble.  The same aperture is then applied to a source detection in every image and a
consistent measure of the light from the same regions of extended objects at different wavelengths is obtained.  While
robust, this method reduces the image quality of all data sets and requires substantial effort in understanding and
processing the imaging pixel data.  The task becomes more challenging when matching data sets over large areas as data
volumes extend to the terabyte regime.  More sophisticated model ``forced'' fitting photometric techniques may provide
better estimates but are even more difficult to execute over large areas.

In an era of rapidly expanding and overlapping imaging surveys, a faster alternative that operates at the catalog level
(instead of the pixel data) is clearly appealing.  This is the motivation for the \synmag{} technique \citep{bundy12}
that we apply here.

Briefly, starting with the available SDSS and UKIDSS catalogs, we cross-match sources using a position tolerance of
0\farcs8 and ``handshaking'' \citep[e.g.,][]{hewett06} to reduce misidentifications.  Handshaking requires that the
nearest UKIDSS source matched to an SDSS object have as its nearest match that same SDSS source.  Thus,
cross-referenced pairs of sources from the two catalogs are uniquely matched.  From the original Coadd parent catalog
with 15,342,585 sources, 3,175,036 (20\%) are matched to a UKIDSS source.  Of the original UKIDSS catalog, matched
sources account for 63\%.  

In the next step, \synmag{}s require that intrinsic, PSF-convolved surface brightness profiles have been fit in at
least one band (we will refer to it as the ``profile band'').  In our case, we use the $r$-band profile fits measured
by the SDSS pipeline, which also delivers the SDSS \texttt {ModelMag}s.  Given the seeing and photometric aperture in a
target band, we use the profile fit to ``predict'' the PSF-matched aperture photometry that would have been observed in
the profile band.  This approach provides a consistent color between these two bands.  We then repeat the process for
the remaining profile--target band combinations, thereby building a full set of matched photometry.  By initially
decomposing the profile fits into Gaussian mixture models
\citep[also see][]{hogg13}, PSF-convolutions can be done analytically, and because all required information is derived
from source catalogs, the technique is extremely fast.  Further details and tests of the methodology are given in
\citet {bundy12}.


\subsection{UKIDSS Total Magnitudes for Extended Sources}\label{total_mags}

Measurements of total magnitudes for extended sources are problematic in the UKIDSS public catalogs.  The WSA ``known
issues'' page\footnote{{\tt http://surveys.roe.ac.uk/wsa/knownIssues.html}} reports several long-standing problems.
Kron radii and magnitudes \citep{kron80} are not reliable in any survey data except for the Ultra Deep Survey.  Other
total flux estimators such as Petrosian \citep{petrosian76} and Hall \citep{hall84} fluxes, as well as their associated radii, are not measured correctly
for blended sources, as we demonstrate below.  Unfortunately, while the overall fraction of UKIDSS sources that are
affected by blends is small (about 1\% among all $K$-band detections in the \cat{}), the fraction rises to 8--10\% for
galaxy samples with $\log{M_*/\msun} > 11.4$ and is magnitude (i.e., redshift) dependent.  For blended sources in the
\cat {}, the solution we adopt is to apply a total magnitude aperture correction to each UKIDSS band that is equal to
the difference between the synthetic aperture $z$-band photometry and the $z$-band \texttt{CModelMag} total flux
estimator. In other words, we will assume that the total flux aperture correction in the $z$-band for these objects
applies to the near-IR bands as well.  Similar assumptions have been tested and applied in other surveys as well
\citep[e.g.,][]{capak07a}.

\begin{figure*}
\plotone{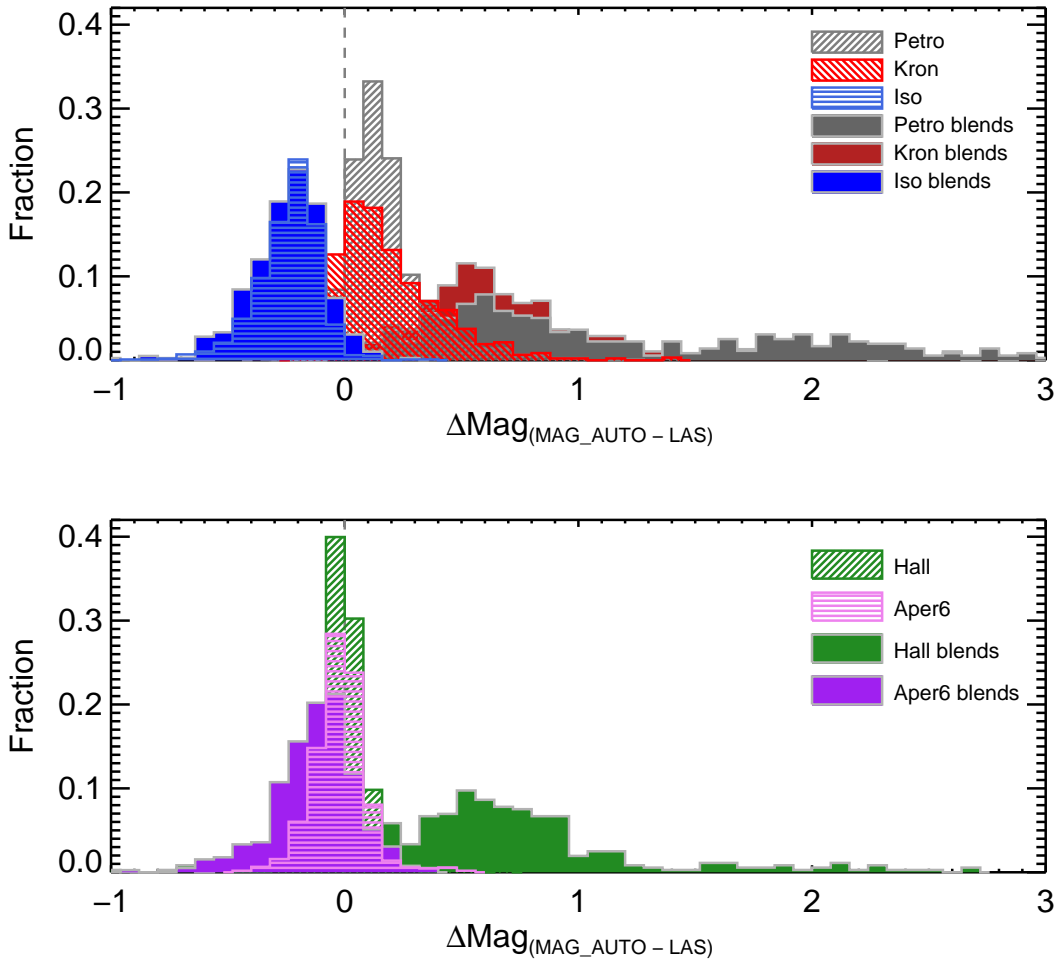}
\caption{Comparisons of various total magnitude estimates from the UKIDSS LAS catalog to MAG\_AUTO
  remeasured with SExtractor.  UKIDSS Petrosian magnitudes (Petro), Kron magnitudes (Kron),
  isophotal magnitudes (Iso), Hall magnitudes (Hall), and corrected ``total'' AperMag6 magnitudes
  (Aper6) are displayed.  Hatched histograms refer to offsets for isolated sources while solid, filled
  histographs show results for deblended sources.  \label{fig:maghist_remeasure}}
\end{figure*}

We begin by demonstrating the amplitude of the UKIDSS total magnitude errors.  We have downloaded nearly 6000
UKIDSS image ``cutouts'' centered on $K$-band bright sources, both isolated and blended and drawn from the \cat{}, each
with a size of $1\arcmin \times 1\arcmin$.  We have run SExtractor \citep{bertin96} on the cutouts and compared its output
magnitudes to various total flux estimators in the UKIDSS DR8 catalog.  The results are reported in Figure
\ref{fig:maghist_remeasure}.  The zeropoint is determined from the keywords stored in the zeroth and first extension
headers in each cutout FITS image, following Equation 1 of
\citet{hill11}:

\begin{equation}
      {\rm ZP} = {\rm ZP_{mag}} + 2.5 \log{t} - {\rm Ext} \times (({\rm AM_1}+{\rm AM_2})/2 - 1) 
\end{equation}

\noindent where ${\rm ZP_{mag}}$ refers to the keyword, MAGZPT, $t$ refers to EXP\_TIME, ${\rm Ext}$
refers to EXTINCT, and ${\rm AM_1}$ and ${\rm AM_2}$ refer to AMSTART and AMEND, the beginning and ending airmass
values of the exposure.  We verify our zeropoint determination by first comparing aperture photometry in the UKIDSS
\texttt{AperMag2} through \texttt{AperMag6} apertures (corresponding to aperture radii of 1.4, 2, 2.8, 4, and 5.7
arcseconds).  Focusing on isolated sources and de-correcting the LAS aperture magnitudes as in Section
\ref{synthetic_mags}, we find that the median LAS photometry is $\sim$0.03 magnitudes brighter in all apertures; we
attribute this to small catalog-level corrections discussed in \citet{hodgkin09}.

Using hatched histograms for isolated sources and solid, filled histograms for blended sources, Figure
\ref{fig:maghist_remeasure} displays blended total photometry errors in the UKIDSS Petrosian magnitudes (Petro),
Kron magnitudes (Kron), Hall magnitudes (Hall), and corrected ``total'' \texttt{AperMag6} magnitudes (Aper6). In all
cases except for isophotal and \texttt{AperMag6} magnitudes, LAS catalog photometry of deblended sources is significantly
brighter than MAG\_AUTO, typically by 0.5--1 magnitude.  Petrosian magnitudes provide the poorest estimates, with offsets
as large as 3 magnitudes.

  \begin{figure}
  \epsscale{1.2}
  \plotone{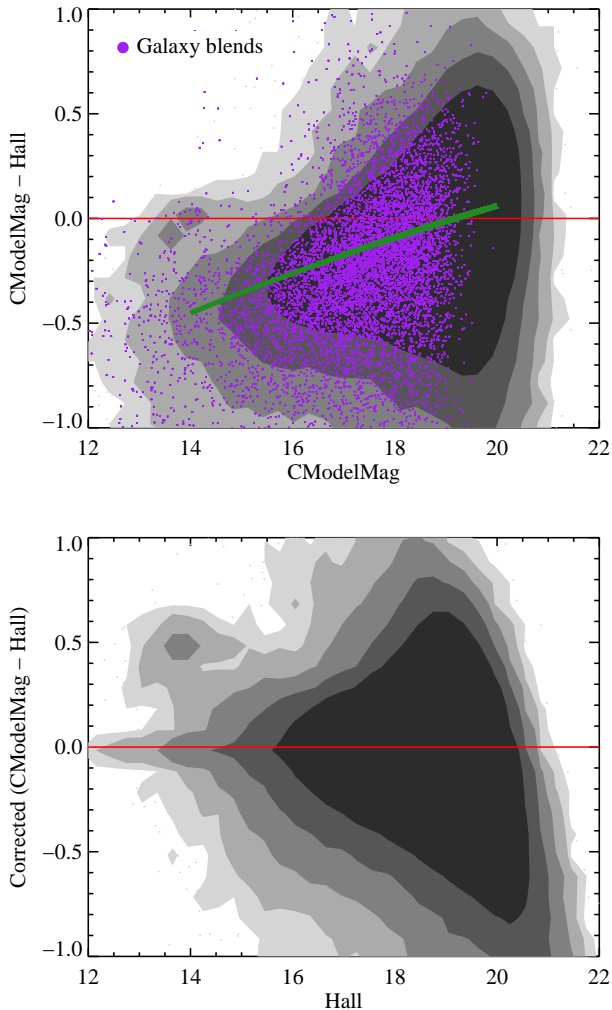}

  \caption{Illustration of the UKIDSS total magnitude estimator adopted in the \cat{}.  For sources
  classified as galaxies we plot $K$-band results in this figure, but the other bands are similar.  Here
  \texttt{CModelMag} refers to the UKIDSS aperture magnitude corrected by the difference between the $z$-band synthetic
  aperture and the $z$-band  \texttt{CModelMag}.  Blended sources are highlighted in purple and fall off sharply beyond
  the detection limit ($\sim$19.5).  In the top panel, the Hall magnitudes are fainter than
  \texttt{CModelMag} for bright galaxies.  The green curve is a fit to this trend.  In the bottom panel, we have
  corrected the Hall mags of isolated sources  using this fit and substituted the \texttt{CModelMag} value for blended
  sources.  This plot of the Hall mag as the dependent variable confirms that there are no biases in the total flux estimator
  above the detection limit.}
  \label{fig:hall}
  \end{figure}

For isolated sources, Hall magnitudes correlate best with MAG\_AUTO but still suffer biases in the case of blended
sources.  The ``total'' magnitudes reported for \texttt{AperMag6} (i.e., with the UKIDSS aperture correction applied)
show the opposite behavior in that blends are {\em fainter} than MAG\_AUTO on average compared to isolated sources.
Even when comparing de-corrected \texttt{Aper6} magnitudes to the corresponding aperture photometry remeasured by
SExtractor, there is a bias such that the reported \texttt{AperMag6} photometry is 0.05 mag fainter for blends.  This
bias decreases as the aperture decreases and nearly disappears for \texttt{Aper4} and smaller apertures.  We speculate
that the reported aperture magnitudes in the UKIDSS catalog have also been adjusted to remove an estimated contribution
from overlapping sources in blended objects, although we could not find UKIDSS documentation to confirm this
conclusion.  The comparison to ``total'' \texttt{AperMag6} values suggests this adjustment introduces a roughly 0.1
magnitude bias towards fainter magnitudes in the large-aperture photometry of blended sources.

The offset for isophotal magnitudes shows no apparent dependence on whether a source is blended.  A simple brightening
of the reported isophotal photometry by $\sim$0.2 magnitudes brings them into agreement with MAG\_AUTO, but the
relatively large scatter in the isophotal offset hints at the obvious problem that isophotal magnitudes cannot account
for surface brightness dimming and will introduce biases as a function of magnitude (or redshift).  

Without a robust total magnitude estimate from UKIDSS for blended sources, we turn to SDSS.  Not only does the SDSS
Coadd photometry deal with blends, it also fits 2D surface brightness profiles to every source. A
recommended\footnote{see \texttt{http://www.sdss.org/dr12/algorithms/magnitudes/\#which\_mags}.} total flux estimator
is the \texttt{CModelMag} which takes the best linear combination of separate de Vaucouleurs and exponential fits in
every SDSS band.  Such profile fitting can account for extended surface brightness profiles that cross under the noise
background, but, constrained by the profile shape assumptions adopted, they can also introduce biases \citep[e.g.,][]
{bernardi13}.

With these caveats in mind, we build a new total magnitude estimate, \texttt{HallTot}, that is referenced to the UKIDSS
Hall magnitudes for non-blended sources and to the SDSS $z$-band \texttt{CModelMag} magnitude for blends.  For blends,
we apply the offset between the $z$-band synthetic aperture magnitude and \texttt{CModelMag$_z$} to the aperture
magnitude (\texttt{AperMag4}) measured for each UKIDSS band.  This choice anchors our UKIDSS total magnitude estimates
to \texttt {CModelMag$_z$}.  However, we would prefer to use the UKIDSS Hall mags for isolated sources, given their
strong performance with respect to the remeasured MAG\_AUTO.  The solution is to adjust the Hall mags to also match
\texttt {CModelMag$_z$} on average.  The process is illustrated in Figure \ref{fig:hall} and delivers blend-resistant
\texttt{HallTot} magnitudes for all four UKIDSS bands.

\section{Star-Galaxy Separation}\label{stargal}

Star-galaxy separation in the \cat{} is more complicated than for typical SDSS imaging data because of problems in the
PSF characterization of the Coadd \citep{annis14, huff14}.  Instead of a classification based on the SDSS \texttt{TYPE}
parameter, we use SDSS shape information as a crude first estimate and then refine the classification using the ($J-K$) vs. ($g-i$) colors to separate galaxies (and quasars) from the stellar
locus.

  \begin{figure}
  \epsscale{1.2}
  \plotone{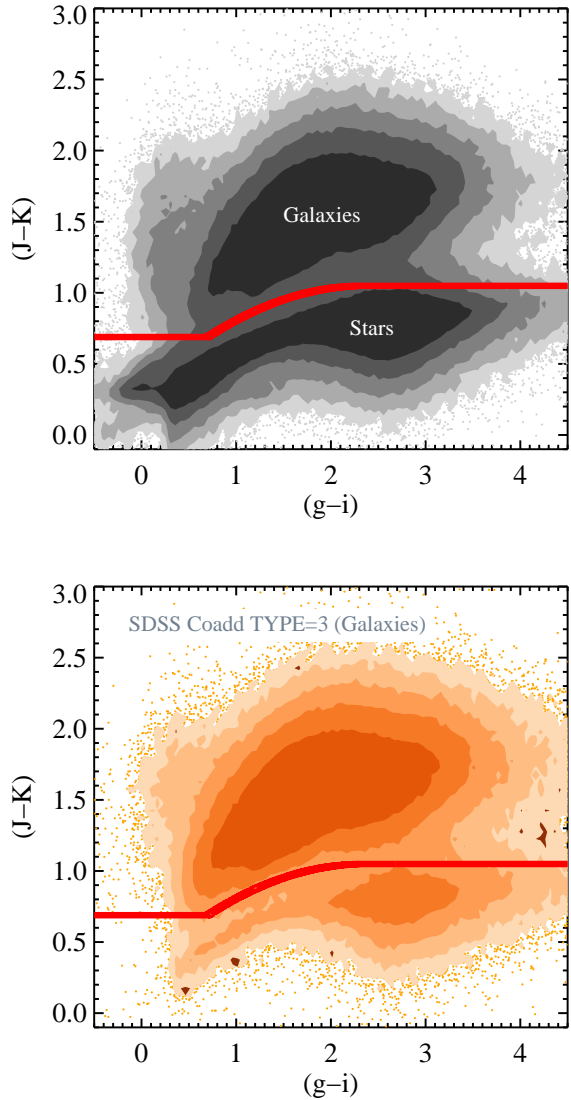}

  \caption{\emph{Top panel}: Color-color criteria for star-galaxy separation adopted in the \cat{}.  Shaded contours
  and data points represent the distribution of sources with $i < 21.5$.  The colors are Galactic extinction corrected.
  \emph{Bottom panel}: Same as above but for SDSS Coadd \texttt{TYPE}$=$3 sources classified as galaxies.  The
  \texttt{TYPE} based galaxy classification is contaminated by a significant fraction ($\sim$10\%) of stars, likely as a
  result of PSF characterization problems in the Coadd.  SDSS Coadd point-sources (\texttt{TYPE}$=$6) do not
  contaminate the color-color based galaxy locus.}

  \label{fig:stargal}

  \end{figure}

Following \citet{baldry10} and \citet{strauss02}, we define the SDSS star-galaxy separation parameter as $\Delta_{\rm
sg} = r_{\rm psf} - r_{\rm model}$, where $r_{\rm psf}$ and $r_{\rm model}$ are the $r$-band PSF and \texttt{ModelMag}
measurements, respectively.  Sources with $\Delta_{\rm sg} > 8$ can be confidently classified as (large) galaxies and
those with $\Delta_{\rm sg} < 0.05$ as point sources.  For the remaining sources, we adopt a modified version of the
($J-K$) distance from the stellar locus defined by \citet{baldry10}:

\begin{equation}
  \Delta_{\rm sg,jk} = (J-K)_{\rm AB} - f_{\rm locus}(g-i)
\end{equation}

\noindent where we correct all colors for Galactic extinction and $f_{\rm locus}(g-i) = -0.523$ for $(g-i) <
0.7$, $f_{\rm locus}(g-i) = -0.1632$ for $(g-i) > 2.3$.  For $0.7 < (g-i) < 2.3$ the locus is defined as:

\begin{equation}
  f_{\rm locus}(g-i) = -0.89 + 0.615*(g-i) + 0.13*(g-i)^2
\end{equation}

We adopt a color cut defined as $\Delta_{\rm sg,jk} = 0.25$ which is displayed in Figure \ref{fig:stargal}.


A check on the completeness and purity of the \cat{} galaxy classification can be made by comparing to the star-galaxy
separator (Leauthaud et al., private communication) based on the overlapping CFHT CS82 $i$-band imaging survey (Kneib
et al., in preparation) which delivers excellent shape-based classifications thanks to a median seeing of
FWHM$=$0\farcs6.  First, we note that different internal classifications using CS82 are impure or incomplete at the
several percent level, representing a quality floor in typical ground-based data sets.  Next, we compare against the
``DW'' CS82 classification which is based on manual divisions made in the size-magnitude plane.  The completeness of
the \cat{} galaxy classification compared to DW is 96\%, with a 1\% level of contamination from stars.  These
numbers may be optimistic because both classifiers can fail.  A visual inspection of SDSS images suggests that
the \cat{} galaxy sample is contaminated by stars (especially binaries) at the few percent level.  

While our classification requires a ($J-K$) color, it performs much better than the Coadd \texttt{TYPE} parameter, used,
for example, by \citet{reis12} to define a galaxy \photoz{} sample.  As shown in the bottom panel of Figure \ref
{fig:stargal}, a Coadd selection with \texttt{TYPE}$=$3 leads to a galaxy sample with at least 10\% contamination from
the stellar locus.

\section{Photometric Redshifts}\label{photoz}

\begin{figure*}
\epsscale{1.2}
\plotone{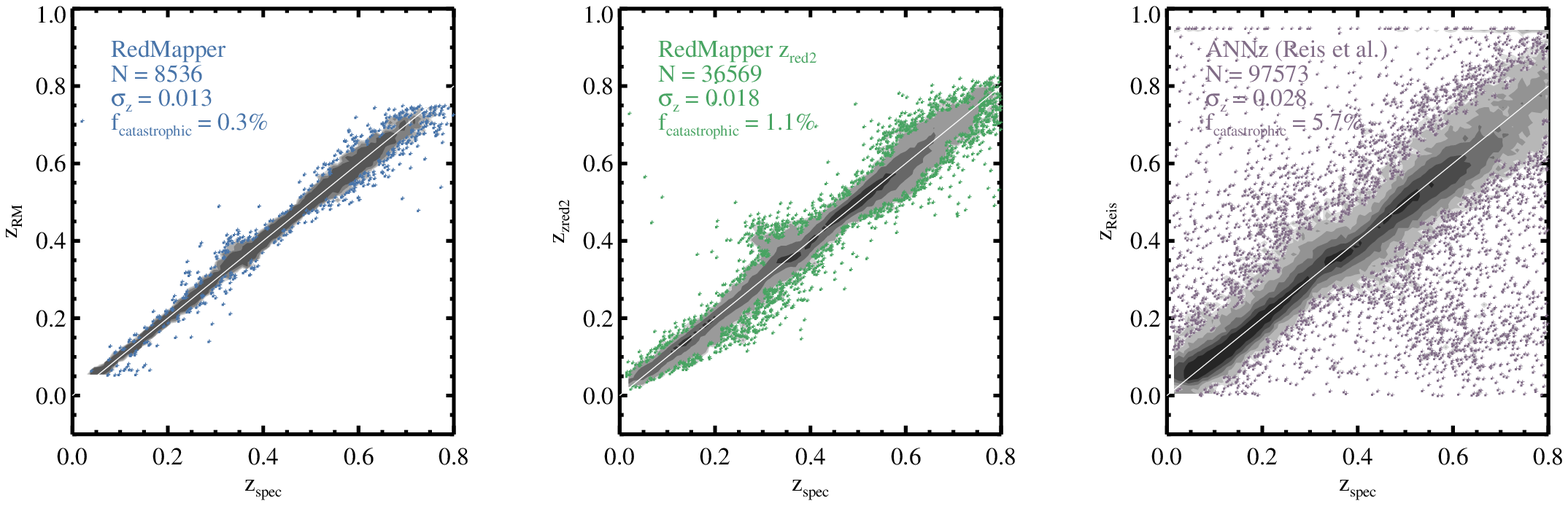}
\caption{Comparisons of three photometric redshift estimators to available spectroscopic redshifts in Stripe 82.  The
  comparison is limited to $i < 22.5$ and $0.01 < z_{\rm spec} < 0.8$.  The left and middle panels are from the
  \redmapper{} project (Rozo et al., in preparation), while the right panel compares neural-network \photozs{} from
  \citet{reis12}. The 3$\sigma$-clipped dispersion is listed in each panel along with the fraction of catastrophic
  outliers defined by $\abs{\Delta z} > 0.1$.  Contours are plotted at high data densities with 0.3 dex logarithmic
  spacing in the left and middle panel and 0.4 dex in the right panel.  The 1-to-1 relation is plotted in each panel as
  a thin light grey line.
  \label{fig:photoz_specz}}
\end{figure*}

\begin{figure}
\epsscale{1.2}
\plotone{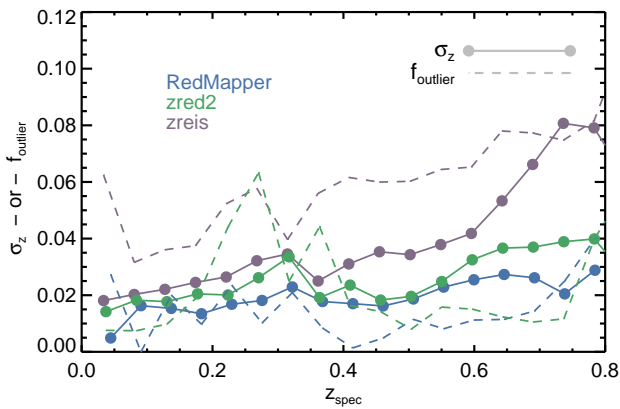}
\caption{Photometric redshift-dependent \photoz{} scatter and outlier
  fractions for the \photoz{} estimators used in this paper as compared to available spec-$z$s in Stripe 82 with $i <
  22.5$ and $0.01 < z_{\rm spec} < 0.8$. The 3$\sigma$-clipped dispersion ($\sigma_z$) is indicated by the solid lines
  and symbols.  The redshift error has not been divided by $(1+z)$.  The dashed lines plot $f_{\rm outlier}$, the fraction of
  \photozs{} that deviate by more than 3$\sigma$; this is a more stringent definition than $f_{\rm catastrophic}$
  sensitive to systematic patterns of scatter that can bias certain redshift bins.  \label{fig:photoz_stats}}
\end{figure}

Despite the large number of spectroscopic redshifts in \cat{} (nearly 73,000 galaxy \speczs{}, or 11\% of galaxies with
$i < 22.5$), reliable photometric redshifts are paramount for building a complete sample in Stripe 82, given that the
majority of the \speczs{} were obtained from SDSS campaigns employing complex selection criteria.  

We have explored a number of photometric redshift estimators and present plots summarizing their performance against
available spectroscopic redshifts in Figures \ref{fig:photoz_specz} and \ref{fig:photoz_stats}.  We break with
tradition and characterize \photoz{} scatter \emph{without} dividing by $(1+z)$  since there is no evidence for such
trends in our data.  We quote $\sigma_z$ as the 3$\sigma$-clipped standard deviation and characterize the outlier
fractions with separate parameters.  We define \emph{catastrophic} outliers as those with $\abs{\Delta z} > 0.1$, where
$\Delta z$ is difference between the photometric and spectroscopic redshift.  Statistical outliers refer to \photozs{}
with $\abs{\Delta z}$ greater than 3$\sigma_z$.

The best \photoz{} performance for massive galaxies in clusters is produced by the red-sequence Matched-filter
Probabilistic Percolation
\citep[\redmapper{},][]{rykoff14}.  For non-cluster galaxies, the best performance is produced by the red-sequence
Matched filter Galaxy Catalog \citep[\redmagic{},][]{rozo15} and neural network results derived in \citet{reis12} using
ANNz
\citep{collister04}.  We use a combination of these estimates to produce the \zbest{} redshift estimator appropriate
for massive galaxies (see Section \ref{sec:zbest}).

We also computed template-based \photozs{} using BPZ \citep{benitez00} and EAZY \citep{brammer08} (see Section
\ref{sub:template_photozs}).  Even after extensive experimentation with various template sets, priors, and zero point
offsets, we could not achieve results that were competitive with ANNz (let alone {\redmapper{}}) for bright galaxies.
The scatter ($\sigma_z$) in
\photoz{}--\specz{} residuals was typically twice as large with the template codes as compared to ANNz, and the
fraction of catastrophic outliers higher by roughly 50\%. However, for faint galaxies ($i \gtrsim 22.5$), template codes
perform somewhat better, likely because of the smaller training sets available in the face of poorer photometry.  We
use template \photozs{} as a check on the neural-network trained redshifts and additionally include them in the \cat{}
with a description below for studies that can benefit from any redshift constraints for the faintest objects in the
catalog.

\subsection{\redmapper{} and \redmagic{} \photozs{}}

The best\footnote{A complete and fair comparison against ANNz would consider just the (primarily red)
galaxies for which \redmapper{} and \redmagic{} \photozs{} are considered trustworthy.} \photozs{} we have compared---with $\sigma_z$
and catastrophic outlier fractions at the 1\% level---are those from the recent \redmapper{} and \redmagic{} catalogs.
 However, these are limited to red cluster galaxies only (\redmapper{}) or red galaxies in general (\redmagic{}).  As described in detail in
\cite{rykoff14}, \redmapper{}'s primary goal is to use the red-sequence method to robustly identify galaxy clusters and
their richness using imaging data sets. By training on available \speczs{},
\redmapper{} iteratively defines a model for the redshift dependent red-sequence that can be used to deliver precise
\photozs{} for red galaxies.



We compare \redmapper\ and \redmagic\ \photozs{} to available \speczs{} in the first two panels of Figure 4. 
\redmapper{} \photozs{} are only available for cluster galaxies, which we define as galaxies in the \redmapper{}
cluster members catalog that have a minimum membership probability of 90\%.  For these galaxies, we assign a
photometric redshift equal to the photometric redshift of the \redmapper{} galaxy cluster.  Roughly 4\% of \cat{}
galaxies with $i<22.5$ fall in this category.   The second estimate we refer to as $z_{\rm red2}$ (or {\sc zred2},
middle panel of Figure 4).  $z_{\rm red2}$ is the starting point for the \redmagic{} photometric redshifts.  It is the
result of applying the empirically derived and redshift-dependent red-sequence template from the cluster finding
algorithm to all red galaxies in the sample, regardless of cluster membership.  Approximately 20\% of \cat{} galaxies
with $i<22.5$ have realizable $z_{\rm red2}$ estimates.  Here, we adopt $z_{\rm red2}$ estimates if the quality
of the \redmapper{} template fit satisfies $\chi^2 \leq 10$ and the galaxy is brighter than $L_*(z)$ luminosity
used by the \redmapper\ and \redmagic\ algorithms.  Our criteria for selecting \redmapper\ and \redmagic\ redshifts
were chosen to balance the number of usable \photozs\ against their quality.  More liberal cuts only marginally
increase the number of available \redmapper\ and \redmagic\ \photozs{}, but significantly decrease their quality.


\subsection{ANNz \photozs{} from Reis et al.}

  \begin{figure*}
  \epsscale{1.2}
  \plotone{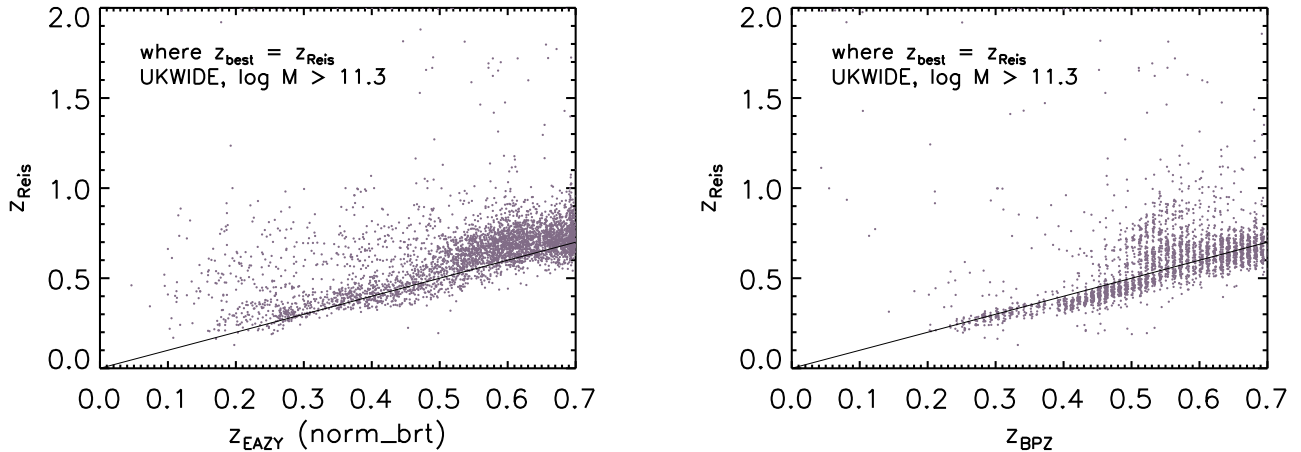}
  \caption{Independent check of neural network ANNz \photozs{}, \zreis{}, against EAZY \photozs{} (left) and BPZ
  \photozs{} (right) for a galaxy sample with no other redshift information available (roughly 30\% of the \ukwide{}
  sample across all redshifts).  Catastrophic problems in \zreis{} resulting from inadequate training sets are not
  evident.  Hints of potential \photoz{} biases at the 0.1 level follow different patterns in the two panels
  and are difficult to confirm because, by definition, no \speczs{} are available for this selection.
  \label{fig:annz_check}}
  \end{figure*}

\citet{reis12} compute \photozs{} using the ANNz neural network code \citep{collister04} applied to $r < 24.5$ sources
from the SDSS Coadd catalog \citep{annis14} with Coadd \texttt{TYPE}$=$3.  As described in Section \ref{stargal}, using
this Coadd \texttt{TYPE} criteria\footnote{Compact galaxies misclassified as point sources would also be missed in the
\citet{reis12} \photoz{} sample.} to select galaxies results in a sample with $\sim$10\% contamination from stars.  A
more accurate star/galaxy separation is presented in Section \ref{stargal} and used to define a complete sample,
\ukwide{}, in Section \ref{sec:ukwide}.  

The \citet{reis12} \photozs{} (we refer to them as \zreis{}) were trained and validated with $\sim$83,000 \speczs{} in
Stripe 82.  They perform well compared to BOSS galaxies despite having been trained
by a more limited \specz{} sample that included only 6,682 BOSS redshifts (Figure \ref{fig:photoz_specz}, right-most
panel).  Comparing to all \speczs{} now available, we find $\sigma_z = 0.028$ and a catastrophic outlier fraction of
5.7\%, in line with the performance reported in \citet{reis12}.

While the training set included additional \speczs{} from deeper surveys, including DEEP2 and VVDS, it is possible that
\zreis{} estimates perform poorly for bright, primarily ``blue'' galaxies that are not well represented in the training
but could be an important contribution to the total massive galaxy population.  We test this possibility in Figure \ref
{fig:annz_check} by comparing \zreis{} \photozs{} to estimates from template codes, \zeazy{} and \zbpz{} (see Section
\ref{sub:template_photozs}).  The comparison is made for sources classified as galaxies in the \cat{} with
$\log{M_*/\msun} > 11.3$ and belonging to the \ukwide{} subsample described below (Section \ref{sec:ukwide}).  We only
plot galaxies for which a \specz{} or \redmapper{} \photoz{} is not available, thus focusing where our best redshift
estimate, \zbest{}, comes from \zreis{}.  About 30\% of \ukwide{} galaxies fall in this category at these masses.

Figure \ref{fig:annz_check} demonstrates that there are no catastrophic problems in \zreis{} resulting from inadequate
training sets for massive galaxies with $z < 0.7$.  While there are hints of potential \photoz{} biases at the 0.1
level, the patterns of these biases are different when comparing to \zeazy{} (left panel) or \zbpz{} (right panel).
Given that all three estimators may have biases, it is difficult to say more about biases in \zreis{} \photozs{} at
this level without additional \speczs{}.

\subsection{Combined \photozs{} for bright galaxies: \zbest{}}\label{sec:zbest}

  \begin{figure*}
  \epsscale{1.0}
  \plotone{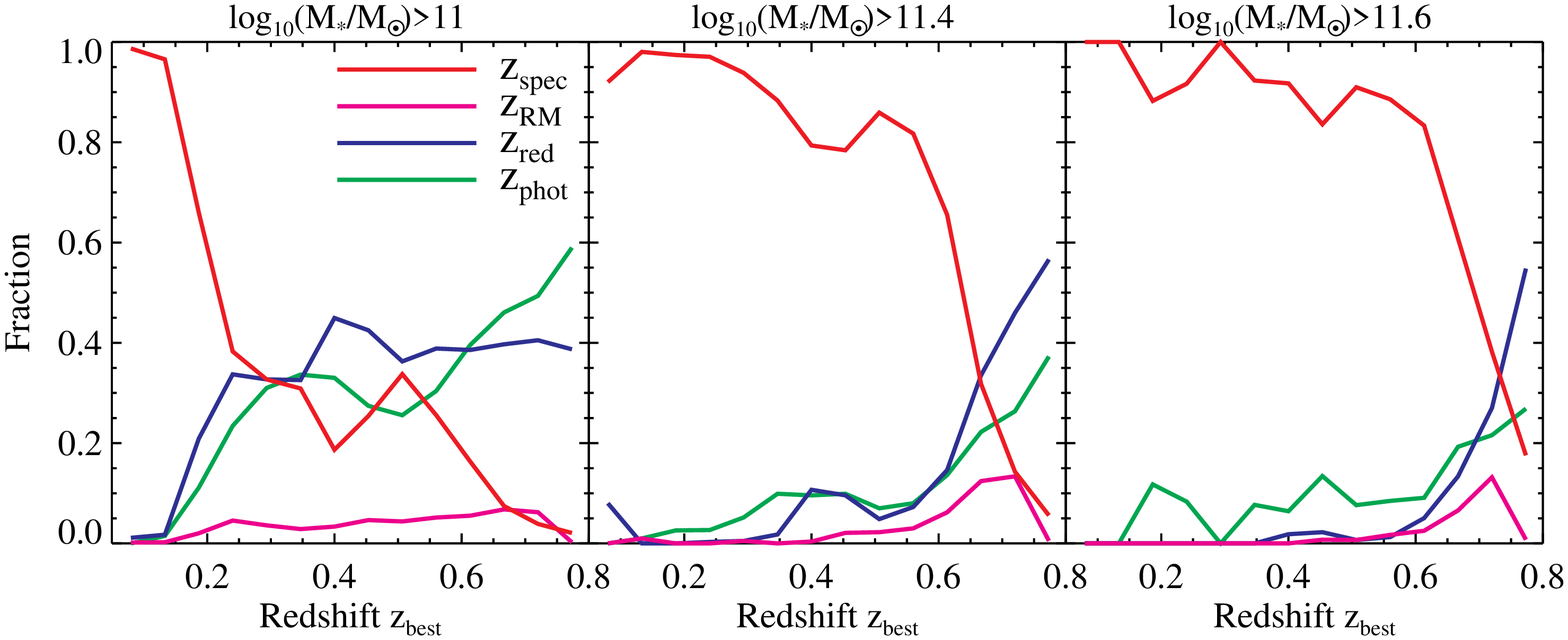}
  \caption{Fractional contribution of different redshift results as a function of $M_*$ and redshift to the ``best'' estimator, \zbest{}, appropriate for
  bright galaxies ($i \lesssim 22.5$) in the \cat{}.  The red line ($z_{\rm spec}$) indicates \specz{} from SDSS.  The
  labels $z_{\rm RM}$ and $z_{\rm red}$ refer to \redmapper{}. ANNz \photozs{} from \citet{reis12} are labeled $z_{\rm
  phot}$ and shown in green.  See text for details.  This figure is reproduced and described further in Paper II.
  \label{fig:zbest_fraction}}
  \end{figure*}

For bright galaxies ($i \lesssim 22.5$), we combine \specz{} and \photoz{} results together into a single ``best''
estimate that we refer to as \zbest{}.  In order of priority, \zbest{} is set
to:

\begin{enumerate}
\item $z_{\rm spec}$: Spectroscopic redshifts passing quality cuts from SDSS (including ``legacy'' and BOSS samples),
VVDS, or DEEP2.
\item $z_{\rm RM}$: \redmapper{} \photoz{} for cluster members.
\item $z_{\rm red2}$: \redmagic{} \photoz{} for red field galaxies.
\item \zreis{}: ANNz \photozs{} from \citet{reis12}.
\end{enumerate}

The fractional contribution of these estimators as a function of both $M_*$ and redshift is shown for the \ukwide{}
sample in Figure \ref{fig:zbest_fraction}.  Paper II studies the SDSS+BOSS spectroscopic completeness evident in this
figure.  Studies using SDSS+BOSS samples to $z \sim 0.7$ are more than 80\% \emph{spectroscopically} complete
at $\log{M_*/\msun} > 11.6$ (right-most panel) and only moderately worse at $\log{M_*/\msun} > 11.4$ (middle panel).




\subsection{Template \photozs{} for faint galaxies} 
\label{sub:template_photozs}

  Template \photozs{} from \EAZY{} \citep{brammer08} and \BPZ{} \citep{benitez00} have also been computed to provide a
  check on \zreis{} (see above) and additional redshift information for faint sources in the \cat{} ($i \gtrsim 22.5$).
  Unlike the \citet{reis12} ANNz catalog, we provide template \photozs{} regardless of star/galaxy classification, in
  part to enable studies of possibly misclassified compact galaxies, but also to allow future improvements in
  separating stars and galaxies.  As we show, the \zreis{},
  \zeazy{}, and \zbpz{} estimates fail in different ways at the faint end of the \cat{} sample.  One advantage of both
  \EAZY{} and \BPZ{} is that some control of the failure rate is possible with cuts on the output quality \texttt{ODDS}
  parameter, if one is willing to limit the usable sample size.  These features make the template \photozs{} useful for
  source galaxies in weak gravitational lensing studies, for example.

  It is well known that template errors in the rest-frame near-IR can lead to degraded \photoz{} performance when
  near-IR data is included \citep[e.g.,][]{brammer08}.  \citet{bundy12} demonstrated acceptable \photoz{} performance
  with near-IR
  \synmag{}s included as a way of validating the \synmag{} methodology.  We confirm that the addition of UKIDSS
  photometry here does not improve the template \photozs{}, but also does not degrade them when the photometry is
  appropriately weighted to take template errors into account.  In what follows, both \zeazy{} and \zbpz{} estimates
  are therefore based solely on the optical $ugriz$ SDSS Coadd (\texttt{ModelMag}) photometry.

  For the \zeazy{} estimates, we built a custom grid of redshift priors as a function of apparent magnitude via
  comparisons to  the SDSS+BOSS \speczs{} at bright magnitudes and the complete VVDS \speczs{} at faint magnitudes.  We
  converted SDSS Coadd inverse hyperbolic sine (asinh) magnitudes (or ``luptitudes'', see \citealt{lupton99}) to
  fluxes and used these as input without any adjustments to the filter band zeropoints.  We
  found the best performance at faint magnitudes using the standard set of \EAZY{} templates.  For bright galaxies ($i
    \sim 20$)---the relevant sample for checking the ANNz \zreis{} redshifts in Figure \ref{fig:annz_check}---we found better results with variants of the \citet{blanton07} templates (BR07).  Aiming to improve outliers
    in the troublesome $z_{\rm spec} \approx 0.35$ range, we combined results from the \EAZY{}-supplied BR07 template
    set with a modified version that removed the most extreme star-forming template.  We further attempted to calibrate
    any remaining \zeazy{} biases and refer to this version of \zeazy{} estimates as ``\texttt{norm\_brt}''.  Compared
    to SDSS+BOSS \speczs{} with no \texttt{ODDS} cut, the \texttt{norm\_brt} sample exhibited a $\sigma_z$ of 0.03 with an outlier fraction of
    18\% and a catastrophic outlier fraction of $\sim$10\%.

  To estimate \zbpz{}, no adjustment to the assumed priors was made, but we did apply the zeropoint offsets that \BPZ{}
  is able to compute for sources with \speczs{}.  We experimented with several different magnitude cuts when
  determining these offsets and also tested several SED template sets.  These experiments led to only small
  improvements in  \photoz{} performance.

\begin{figure*}
  \epsscale{1.2}
  \plotone{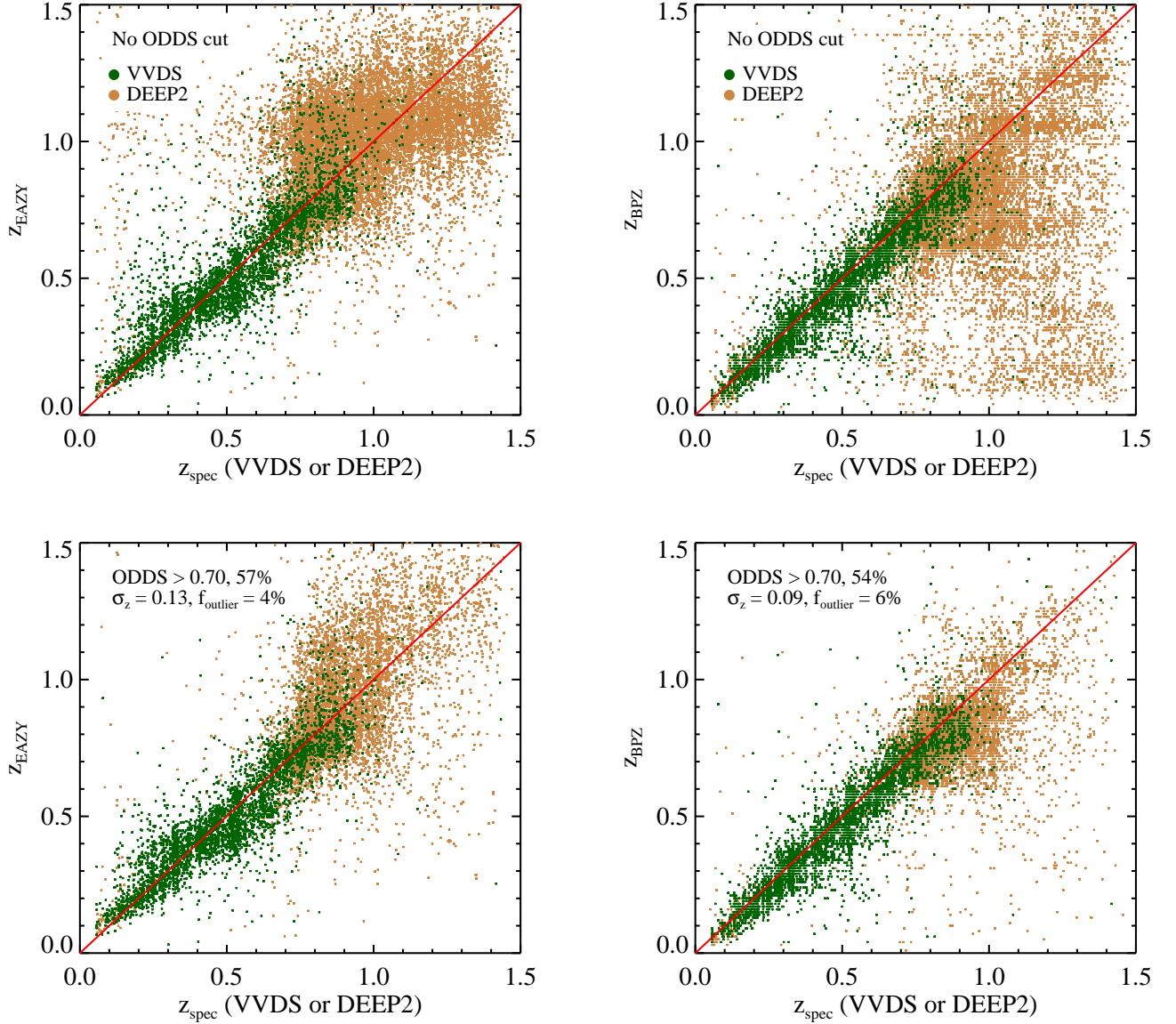}
  \caption{Photometric redshift quality for the \emph{faintest} galaxies in the \cat{}.  Results
  using the template-based \EAZY{} (left panels) and \BPZ{} (right panels) \photoz{} codes are compared to VVDS and DEEP2
  \speczs{} as indicated.  VVDS is a magnitude limited survey with $i \lesssim 22.5$, while DEEP2 targets were
  color-selected to have $z \gtrsim 0.7$ and typically have magnitudes of $i \gtrsim 22.5$ which therefore reach or
  exceed the
  \cat{} detection limits.  Catastrophic outliers display significantly different behavior between the two codes.
  Given sparse training sets at faint magnitudes, the \texttt{ODDS} parameter output from \EAZY{} and \BPZ{} provides
  an advantage over neural-network
  \photozs{} for selecting better performing \photoz{} samples.  For example, the lower panels show improved  results
  for \texttt{ODDS}$>$0.7.  The percentage of the original sample that satisfies the \texttt{ODDS} cut is labeled in
  each panel as are metrics characterizing the resulting \photoz{} quality.
  \label{fig:zfaint_template}}
  \end{figure*}

\begin{figure}
  \epsscale{1.2}
  \plotone{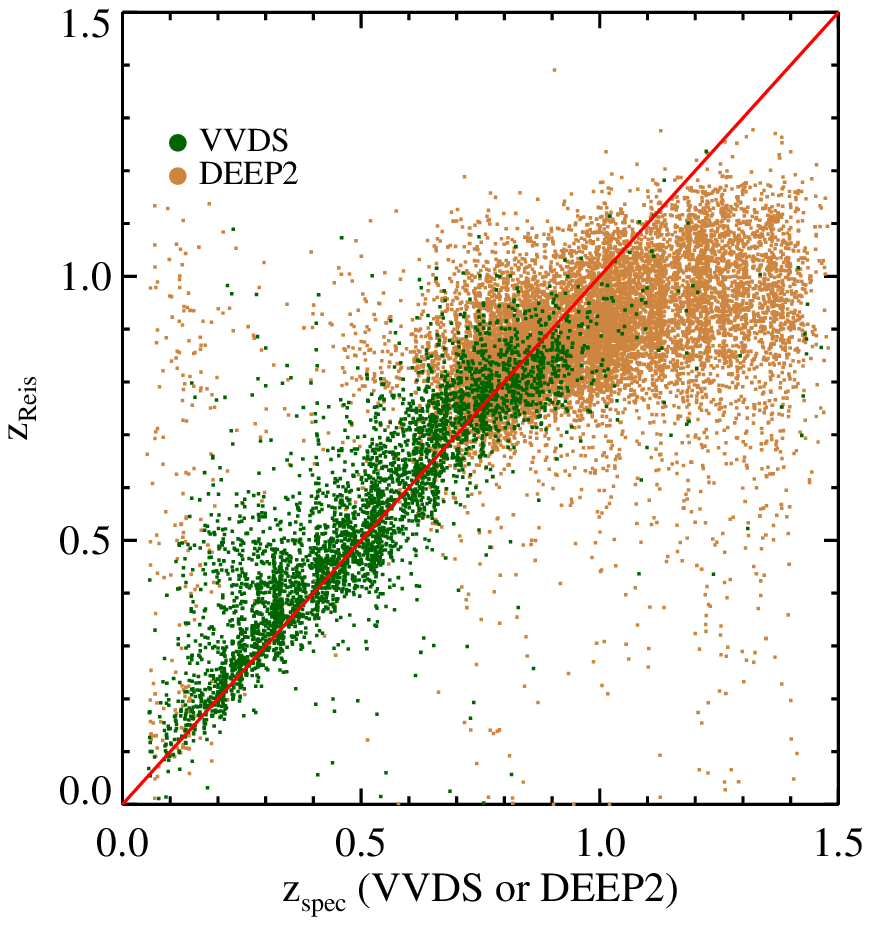}
  \caption{Performance of the \zreis{} neural-network \photozs{} when compared to the VVDS and DEEP2 samples used in Figure
  \ref{fig:zfaint_template}.  The \zreis{} estimates were trained in part on these \speczs{} but show larger scatter
  ($\sigma_z)$ than the template fitting \photoz{} estimates compared in Figure \ref{fig:zfaint_template}.  It is also
  more difficult to prune high-performing subsets of neural-network estimates.
  \label{fig:zfaint_annz}}
  \end{figure}

  Given the superior performance of the \redmapper{} and ANNz redshifts described above for bright galaxies, we turn
  now to the quality of \zeazy{} and \zbpz{} estimates for the faintest galaxies detected in the \cat{}.  Here \zeazy{}
  is derived using the standard \EAZY{} template set.  Figure \ref{fig:zfaint_template} presents comparisons to VVDS
  and DEEP2
  \speczs{}.   VVDS is a magnitude limited survey with $i \lesssim 22.5$, while DEEP2 targets were
  color-selected to have $z \gtrsim 0.7$ and typically have magnitudes of $i \gtrsim 22.5$ which therefore reach or
  exceed the
  \cat{} detection limits.   Catastrophic outliers in the \zeazy{} estimates tend to congregate near \zeazy{}$ \sim
  1.1$, likely reflecting the custom priors we have adopted.  The \zbpz{} outliers are more distributed and tend to
  fall below the true redshift.  Particular science cases may prefer one or the other failure mode.  In addition, cuts
  on the \texttt{ODDS} \photoz{} quality parameter allows cleaner \photoz{} samples to be obtained (lower panels).  For
  comparison, the \zreis{} estimates for the DEEP2 sample are shown in Figure \ref{fig:zfaint_annz}.

\section{Survey Geometry}\label{geometry}

A precise measurement of the total valid solid angle over which the \cat{} sample was drawn is obviously critical to any
derivation of source densities.  Unfortunately, this measurement is complicated by the relatively large number of intersecting
data sets: the Stripe 82 Coadd imaging, the UKIDSS photometry in four separate bands, and the BOSS spectroscopy.  Not only must we
define the geometric footprint of each data set but also the regions that should be rejected for various reasons, including image
artificats and bright stars.

To do so, we have relied heavily on the \mangle{} software package \citep{swanson08}\footnote{\tt
  http://space.mit.edu/\~molly/mangle} supplemented with both custom built routines and IDL procedures bundled with the SDSS
\idlutils{}\footnote{\texttt{http://www.sdss.org/dr12/software/idlutils}} package.  With \mangle{} it is possible to create, manipulate, and determine the intersections and overlap of
spherical polygons on the sky.  We refer to a set of such polygons as a
``mask,'' which can represent a survey's on-sky footprint or a series of
``holes'' where data should not be considered.  Due to the presence of a bug
in combining polygon weights in previous \mangle{} versions, we used a
development version downloaded in June 2014 from the \mangle{} GitHub
repository\footnote{\tt https://github.com/mollyswanson/mangle}.

\begin{figure}
\epsscale{1.2}
\plotone{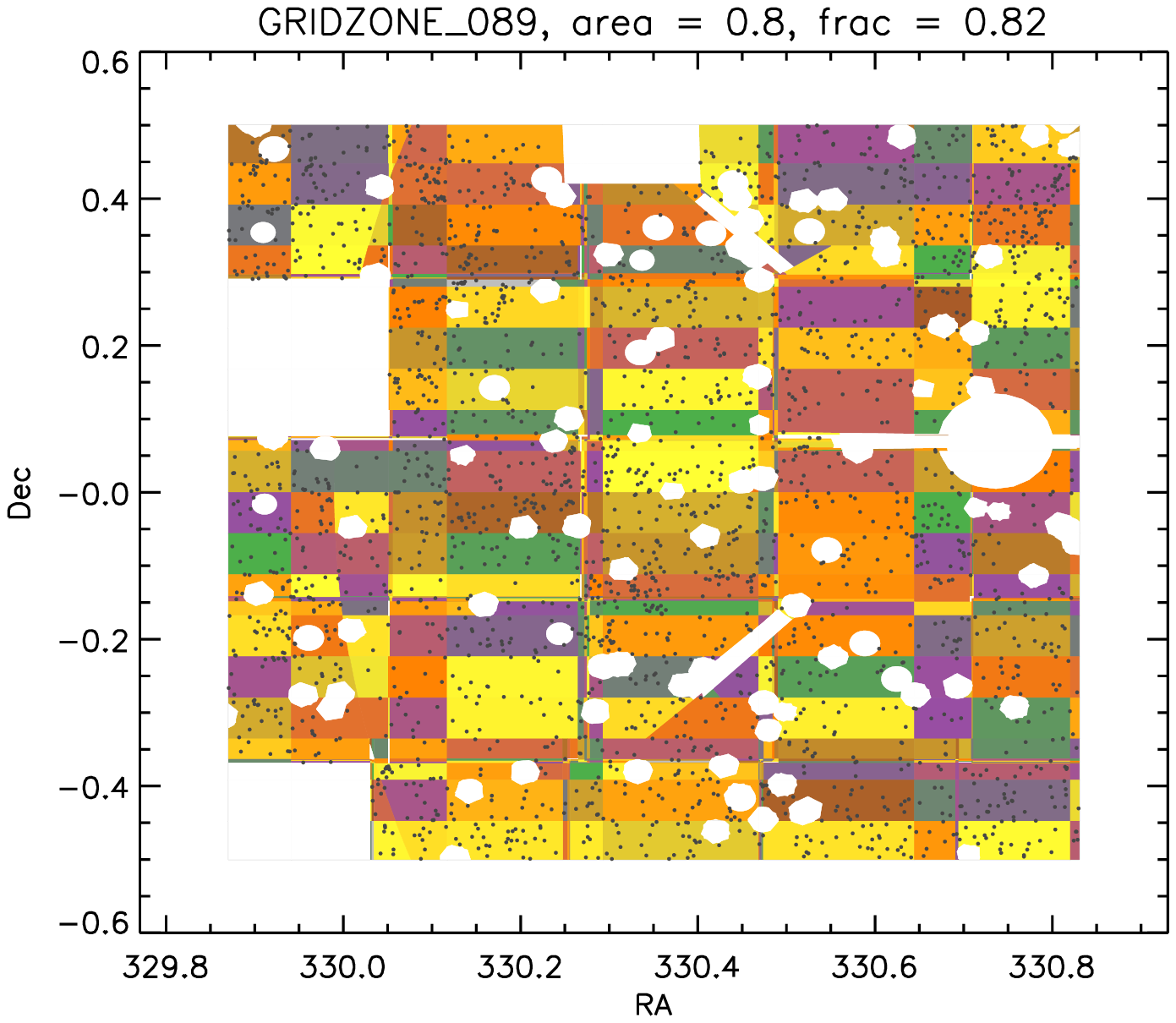}
\caption{A small portion of \cat{} illustrating the union of acceptable survey polygons, each shaded with different
colors for visual clarity.  White regions correspond to rejection masks for stars, missing data, imaging artifacts, and
  BOSS targeting limitations.  The positions of galaxies from the \ukwide{} subsample (see Section \ref{sec:ukwide})
  are overplotted. \label{fig:polygons}}
\end{figure}

All \mangle{} polygon files in \cat{} are ``pixelized'' to a fixed resolution of $r = 11$ and ``snapped.''  We adopt
angle tolerances following the DES collaboration (Aur\'elien Benoit-L\'evy, private communication) with cap tolerance
variables, ``a,'' ``b,'' and ``t,'' set to 0\farcs0027 and the intersection angle tolerance, ``m,'' set to
$10^{-8}$\arcsec.  From a visual inspection of output polygon masks, we estimate that these \mangle{} settings yield
total area estimates that are accurate at the 2\% level.  A small $\sim$0.8 deg$^2$ portion of the final
polygon file, with source detections overplotted, is presented in Figure \ref{fig:polygons}.

\subsection{BOSS Survey Masks}

Because the BOSS team has used \mangle{} tools extensively, several useful masks defining the BOSS survey already exist
and are publicly available.  We used masks described on the BOSS DR9 Large Scale Structure page\footnote{\tt
https://www.sdss3.org/dr9/tutorials/lss\_galaxy.php} on the SDSS-III website.   There is little to no difference in the
BOSS DR10 masks because no new observations were obtained in Stripe 82 after DR9.  Because the BOSS footprint overlaps
almost entirely with \cat{}, applying the BOSS masks when selecting the final \cat{} sample enables studies of the
completeness of BOSS samples (\comppaper{}) and reduces the total area by
less than 10\%, almost entirely as a result of the BOSS collision priority rejection mask (see Section \ref{rejmasks}).

The BOSS survey footprint (acceptance mask) is defined in the \mangle{} polygon file, {\tt boss\_geometry\_2011\_06\_10.ply}.  We
also make use of several rejection masks.  Beyond the bright star rejection mask, {\tt bright\_star\_mask\_pix.ply}, the
centerpost mask, {\tt centerpost\_mask.ply}, removes a small region associated with the center of each BOSS plate where
spectroscopic targets could not be observed.  The collision priority mask, {\tt collision\_priority\_mask.ply}, accounts for small
zones around BOSS galaxy targets that may have been de-prioritized in favor of other source classes (such as quasars). 
Before application to \cat{}, all BOSS masks are trimmed to the Stripe 82 region.

\subsection{UKIDSS footprint masks}

We derive the areal geometry of the four filter bands in the UKIDSS LAS photometry in Stripe 82 using custom software and make the results publicly available on the
\cat{} website.  We first define a template for the 4-detector WFCAM imaging footprint following the field layout information on the UKIRT
website\footnote{\tt http://www.jach.hawaii.edu/UKIRT/instruments/wfcam/user\_guide/description.html}, and then
identify all catalog sources associated with a specific WFCAM {\it multiframe}.  \emph{Multiframes} can overlap, miss
data
from certain detectors, or be truncated, for example from spilling over the edge of the SDSS Coadd region that defined
the positional boundaries in the UKDISS catalog query.

For these reasons, we use the catalog source positions to reverse engineer each \emph{multiframe}'s sky position and determine the
geometry of its acceptable imaging.  We construct heavily smoothed histograms of the celestial coordinates of associated sources
in order to identify the gaps that define the WFCAM footprint.  We then adjust the footprint template to match, deriving spherical
rectangles that correspond to each detector.  Where \emph{multiframes} are truncated we use this gap information (if a
gap exists) and the most extreme source positions to define the maximum extent of each detector.  The resulting ``framemask'' is
assigned a ``weight'' equal to the derived AB magnitude imaging depth discussed in Section \ref{ukidss_depth} and is converted into the
\mangle{} polygon format.  The pixelized and snapped UKIDSS footprint masks have file names of the form, {\tt
  mangle\_framemask\_\{y,j,h,k\}\_v???.ply}, where {\tt v???} represents a version number.

\subsection{Rejection masks}\label{rejmasks}

Even in regions with valid UKIDSS or SDSS Coadd photometry, it is important to mask regions around bright stars and
associated diffraction spikes, satellite trails, and other image artifacts.  Ideally, one would define masks by
inspecting actual imaging data, but for wide-field surveys with many bands, this is a challenging data management
problem.  Instead, we inspect the spatial distribution of cataloged sources across the survey region, distinguishing
valid photometric sources from those flagged because of a problem identified by the photometric pipeline.
Problematic regions can be identified easily by eye because they exhibit an overdensity of ``bad'' sources or trace an
unphysical geometric pattern (such as a diffraction spike).

We developed a software package that takes a catalog of source positions as input and allows the user to scan through
the survey footprint.  Bad areas can be traced interactively.  The user can draw either a circle by defining the center
with one click and the radius with a second, or a generic polygon by clicking a set of points to define the polygon
vertices.  These shapes are saved, over-plotted, and can be deleted and redrawn.  When finished, the
software converts the stored shapes into a
\mangle{} polygon file that can be used as a rejection mask to define the final survey footprint.

This technique was used to define the rejection masks for all four UKIDSS bands as well as the SDSS Coadd catalog.  For
each UKIDSS band, the amount of rejected area is $\sim$3 deg$^2$.  Masking of the SDSS Coadd accounts for 1.2 deg$^2$.
Most bright stars were already masked by the BOSS bright star rejection mask (4.3 deg$^2$ over \cat{}),
but several were missed or were more of a problem in the UKIDSS bands.  In the \cat{} region, the BOSS collision
priority mask sums to 16.8 deg$^2$ and the centerpost rejection mask amounts to 0.1 deg$^2$.  The rejection masks are
available on the \cat{} website.

\section{The \ukwide{} sample}\label{sec:ukwide}\label{sample}

With the parent SDSS Coadd and UKIDSS data sets defined above, we select a galaxy sample optimized to have a well
characterized $M_*$ completeness limit as wide an area as possible in the \cat{}.  We refer to this sample as
``\ukwide{}.''  Forming the basis for the number density evolution and BOSS completeness studies described in
\comppaper{}, \ukwide{} contains 517,714 galaxies with matched Coadd+UKIDSS photometry.  It covers \area{} deg$^2$, and
is complete above $\log M_*/\msun \approx 11.2$ at $z = 0.7$.  Data tables corresponding to the \ukwide{} sample are
made publicly available as described in the Appendix.

The \ukwide{} sample applies several initial cuts.  Galaxies are separated from stars as described in Section
\ref{stargal}.  We then remove sources located in any of the rejection masks, accounting for bad regions in all four
bands of the UKIDSS photometry and the SDSS Coadd imaging, as well as bright stars and zones where BOSS spectroscopy
was not possible or incomplete (see Section \ref{geometry} for more details).  We also require the \ukwide{} galaxies
to fall in the BOSS DR9 acceptance mask and pass cuts on the UKIDSS \texttt{ppErrBits} error flag.

We must also define the acceptable UKIDSS imaging depths for the \ukwide{} sample.  In AB magnitudes, we choose
5$\sigma$ magnitude limits of $[20.32, 19.99, 19.56, 19.41]$ for the four bands, $YJHK$ (see \ref{ukidss_depth}).
\ukwide{} contains only those sources drawn from UKIDSS imaging with depths greater than these values.  The motivation
for these limits is discussed further below.

\subsection{$M_*$ Completeness Considerations}\label{sec:completeness}

In constructing the \ukwide{} sample, we have adopted a selection that maximizes the available area with relatively uniform
completeness in stellar mass.  A conservative estimate for this redshift-dependent $M_*$ completeness limit in any filter can be
obtained by considering the theoretical range of galaxy stellar mass-to-light ratios ($M_*/L$) present at each redshift.  If we
choose a reasonable representative SED for the maximum $M_*/L$ possible (e.g., a stellar population that formed quickly in a burst of
star formation at $z \sim 5$) and compare it against the limiting magnitude in the filter band, we can estimate the minimum $M_*$
of all galaxies at that redshift with brighter observed magnitudes.

\begin{figure}
\epsscale{1.2}
\plotone{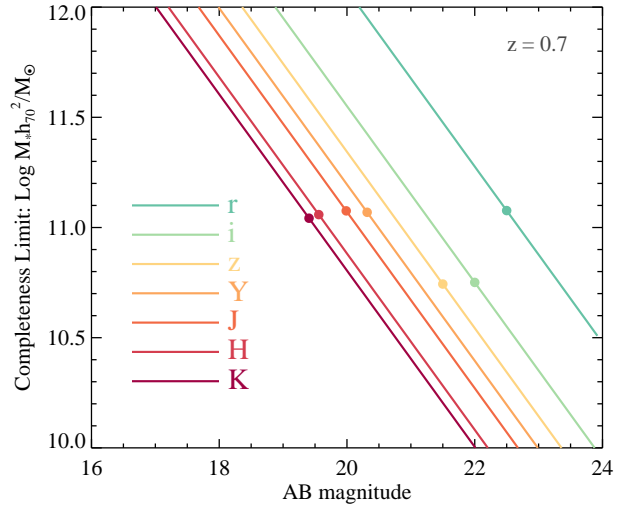}
\caption{The estimated $M_*$ completeness limit as a function of the limiting depth in relevant \cat{} filter bands at
  a specific redshift.  Solid symbols indicate the locations of detection limits in each plotted band.  The SDSS Coadd
  $r$-band depth is comparable to the UKIDSS depths in terms of $M_*$ completeness. \label{fig:masslimit-mag}}
\end{figure}

For a reference redshift, we choose $z = 0.7$, which roughly corresponds to the redshift limit of the BOSS spectroscopy
and the point where the available photometric redshifts become increasingly imprecise.  Figure \ref{fig:masslimit-mag}
displays the $M_*$ limits described above as a function of the magnitude in several bands at this redshift.  We focus
on the $rizYJHK$ as they provide the most critical constraints on $M_*/L$ and $M_*$ at these redshifts in the
\cat{} data set. The ideal magnitude limits in each band at some desired $M_*$ limit are simply given by the
intersection of the plotted lines and that $M_*$ limit.  The observed $r$-band limit is significantly fainter
than the other bands, a result of the 4000\AA\ break at this redshift and the red colors defining the extreme-$M_*/L$
SED template.

The SDSS Coadd $riz$ 90\% completeness depths are indicated by the solid symbols on the corresponding $riz$ lines.  The
$r$-band depth limits the completeness of any \cat{} sample to $\log M_*/\msun
\approx 11.1$.  The redder bands can afford brighter magnitude limits (in AB units) and still have deeper $M_*$ limits at $z =
0.7$.  By roughly matching the $M_*$ limit defined by the $r$-band, we can define ideal magnitude limits (5$\sigma$)
for the UKIDSS bands.  As shown again by the solid symbols for the $YJHK$ bands, our final adopted limits are indeed
close to $\log M_*/\msun \approx 11.1$ at $z = 0.7$.  In practice, the precise numbers ($[20.32, 19.99, 19.56, 19.41]$
for $YJHK$) were matched to a limit at somewhat lower redshift before testing revealed the fidelity of \cat{} to
redshifts approaching 0.7.

\begin{figure}
\epsscale{1.2}
\plotone{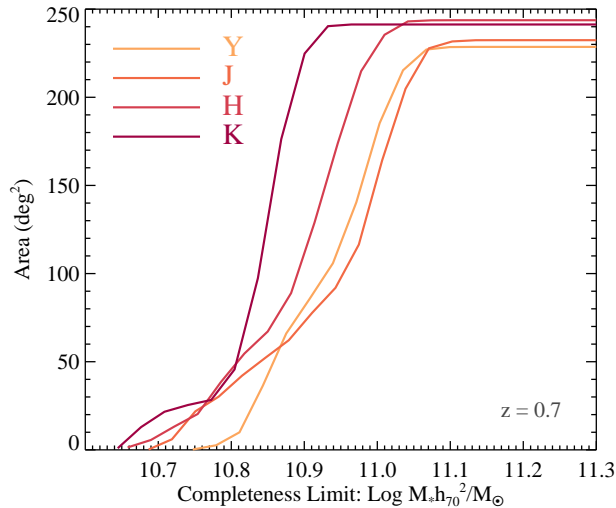}
\caption{We plot the area covered in each UKIDSS filter band as a function of the limiting depth expressed in terms of
  the $M_*$ completeness at a specific redshift.  Assuming that all other bands are sufficiently deep, the completeness
  limit corresponds to the faintest theoretical galaxy with a maximum reasonable $M/L$ that is detected at 5$\sigma$ in
  the relevant band.   \label{fig:area-mass}}
\end{figure}

Because the UKIDSS imaging depth is not uniform across Stripe 82, a final consideration for the adopted magnitude
limits is the corresponding amount of area available.  Working in terms of the $M_*$ limit defined at the 5$\sigma$
detection threshold, Figure \ref{fig:area-mass} displays the cumulative distribution of area imaged to depths greater
than that limit for each UKIDSS band, again referenced to $z = 0.7$.  A limit of $\log M_*/\msun \approx 11.1$ at this
redshift makes use of all the UKIDSS area available.

We quote the final $M_*$ completeness limit for \ukwide{} at $\log M_*/\msun \approx 11.2$.  This is a conservative choice that
acknowledges potential systematics in the way we estimate the UKIDSS and SDSS Coadd depths, as well as catastrophic errors that
may occur in matched photometry and SED fitting when using sources measured at the 5$\sigma$ limit.  Effectively, a limit of $\log
M_*/\msun \approx 11.2$ corresponds to a brightening of our nominal magnitude limits by 0.25 magnitudes.

\section{Derived Galaxy Properties}\label{properties}

\subsection{The birth parameter, \birth{}}\label{birth}

\begin{figure}
\epsscale{1.2}
\plotone{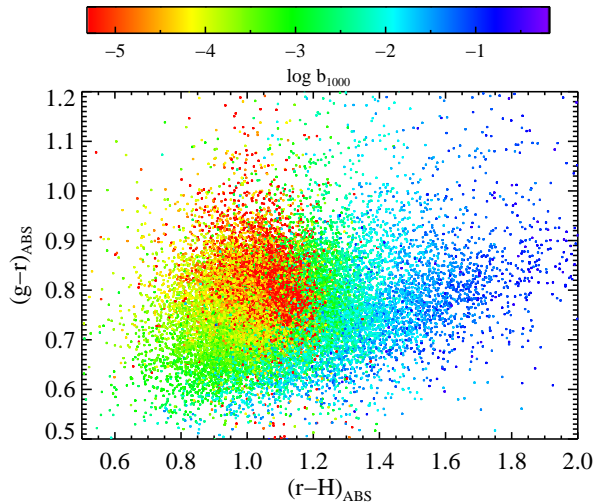}
\caption{Correlation between the \birth{} birth parameter, a proxy for recent star formation, and position in a restframe optical-near-IR color-color diagram
for a \ukwide{} sample with $0.3 < z_{\rm best} < 0.6$ and $\log{M_*/\msun} > 11.2$. 
The value of $\log{b_{1000}}$ is indicated by the associated color, as labeled.  The most quiescent galaxies with the
lowest \birth{} parameters (colored red) are clustered at red ($g-r$) colors but relatively blue ($r-H$) colors, as
expected \citep[e.g.,][]{williams09}.  Galaxies with more recent star formation (higher \birth{}, bluer colors) span a
more extended range in this diagram.  Dusty star-formers continue upwards into the optical red sequence, but can
be distinguished by their redder ($r-H$) colors.}

\label{fig:birth}
\end{figure}

The presence of recent star formation among galaxies in imaging data sets is often inferred from a rest-frame color
that straddles the 4000 \AA\ break.  A major benefit of near-IR data is the ability to use color-color diagrams to
distinguish red galaxies with underlying but dusty star formation from truly quiescent systems
\citep[e.g.,][]{williams09}.  Here we exploit the full optical through near-IR photometry in the \cat{} to estimate a
proxy for recent star formation that accounts for dust.  We compress the 2D information in a color-color diagram into a single measurement by fitting the full SEDs available in the \cat{} with the
\kcorrect{} package \citep{blanton07}.  \kcorrect{} provides an estimate of the birth parameter \birth{}, defined
as the ratio of the average star formation rate within the previous 1 Gyr to the star formation rate averaged over the
galaxy's history.  As with $M_*$ estimates, \birth{} depends on the assumed models and priors used to the fit the
observed SEDs.  Figure \ref{fig:birth} illustrates the correlation between \birth{} and location in the ($g-r$) versus
($r-H$) diagram.  A tight grouping of galaxies with very low \birth{} values is evident as is a more extended
star-forming sequence that includes dusty systems that would otherwise be located on the (optical) red sequence. Visual
inspection of the CS82 imaging confirms that these galaxies are predominantly disk-like or disturbed.  The use of a
near-IR based
\birth{} parameter allows them to be easily distinguished from truly quiescent systems.

In the \cat{}, galaxies at $z\sim0.55$ and $M_*\sim11.5$ have \birth{} values as high as 0.7, suggesting an
occasional high rate of recent star formation, but the vast majority are peaked near $b_{1000} = 0$ as expected for an
old, passively evolving stellar population.

\subsection{Stellar Mass Estimates}

A key motivation of \cat{} is providing near-IR photometry for robust $M_*$ estimates since the range of possible
stellar $M/L$ ratios for stellar populations decreases at near-IR wavelengths.  Among the most massive galaxies with $z
< 1$, the expectation is that the majority have little star formation and are passively evolving.  Thus, $M_*$
estimates based on optical photometry alone may be sufficient \citep[e.g.,][]{pforr12}.  Still, observed $i$-band
corresponds to restframe $g$-band at $z \sim 0.7$, and even the observed $z$-band, which is typically shallower than
$r$ or $i$ in single-epoch SDSS imaging, falls blueward of restframe $r$-band.  Because young stars significantly bias
$M_*$ estimates limited to blue restframe wavelengths, the fact that nearly 40\% of the CMASS sample is intrinsically
associated with the blue cloud \citep{montero-dorta14} and the evidence for an increase with redshift in the fraction
of CMASS galaxies with spectra indicating recent star formation \citep{chen12} motivate restframe near-IR photometry as
an important ingredient for checking evolutionary results\footnote{Paper III presents tests that, while revealing
biases in optical $M_*$ estimates for some galaxies, show that the choice of stellar synthesis modeling and priors has
a larger effect on evolutionary signals for $\log{M_*/\msun} > 11.3$ and $z < 0.7$ than the use of optical-only $M_*$
estimates.} based on estimates of $M_*$.

\subsubsection{\cat{} fiducial $M_*$ estimates}\label{KBmass}

The fiducial $M_*$ estimates derived for the \cat{} use the Bayesian code initially presented in \citet{bundy06} and
used in \citet{bundy10}.  Paper III presents additional mass estimates testing various priors with \isedfit{}
\citep{moustakas13} based on the \cat{} photometry for the \ukwide{} sample.  The observed SED of each galaxy is
compared to a grid of 13,440 models from the \citet{bruzual03} (BC03) population synthesis code that define a set of
priors spanning a (randomized) range of metallicities, star formation histories (parameterized as exponentials), ages,
and dust content. Ages are restricted to less than the cosmic age at each redshift and no bursts are included.  A
Chabrier IMF
\citep{chabrier03}, $\Omega_M$=0.3, $\Omega_{\Lambda}$=0.7, and a Hubble constant of 70 km s$^{-1}$ Mpc$^{-1}$ were
adopted.

At each grid point, the $M_*/L$ ratio in the reddest observed band, the $M_*$ value inferred from multiplying $M_*/L$
by the luminosity in this band, and the likelihood that the model matches the observed SED are stored. This likelihood
is marginalized over the grid, giving an estimate of the stellar mass probability distribution\footnote{We assume the
prior grid adequately samples the parameter space of the posterior.}.  We take the median as the final estimate of
$M_*$, which counts the ``current'' mass in stars and stellar remnants. The 68\% width of the distribution provides an
uncertainty value which is typically $\sim$0.1 dex.

    \begin{figure*}
    \epsscale{1.2}
    \plotone{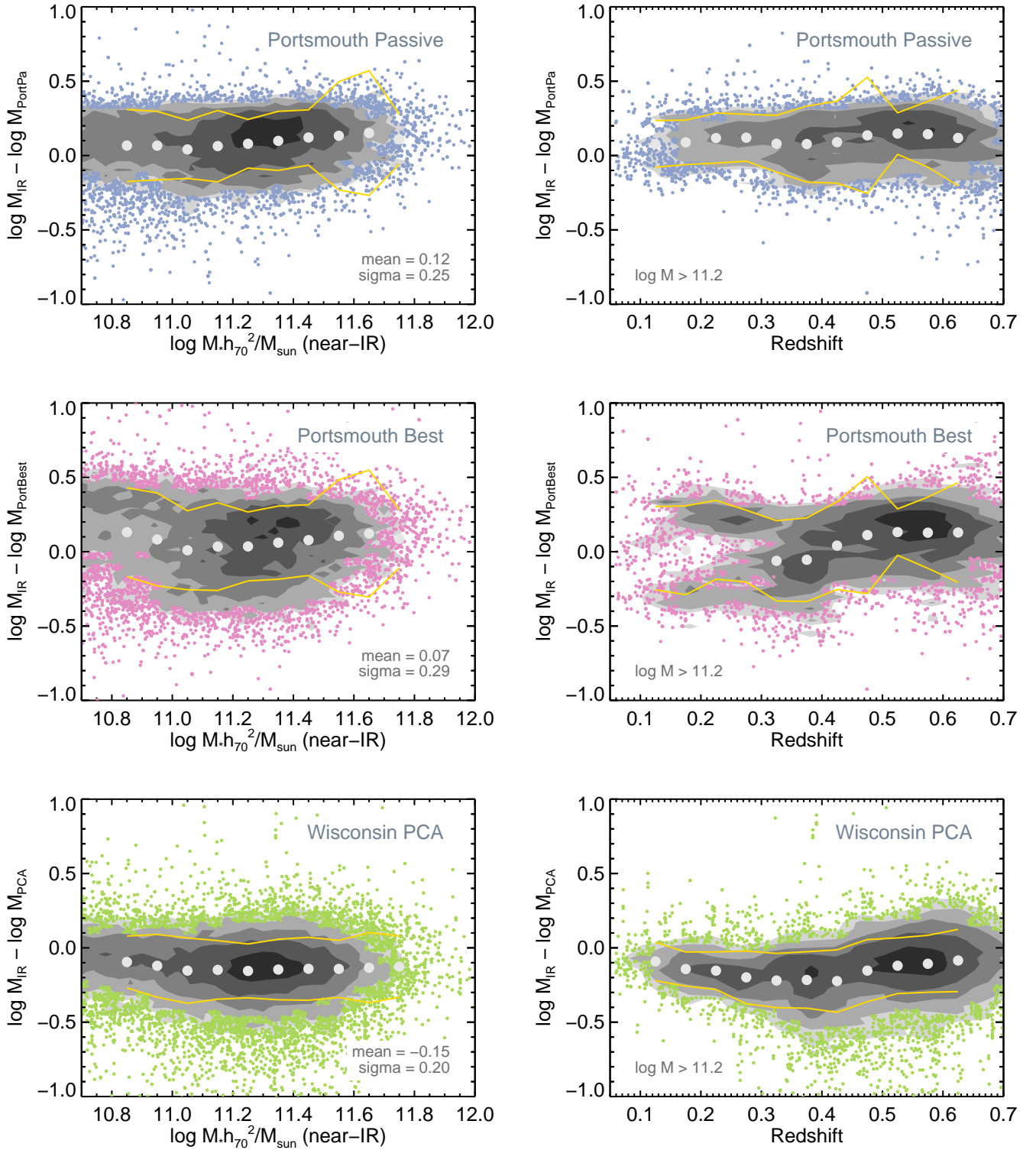}

      \caption{Comparison of public BOSS $M_*$ estimates from \citet{maraston13} and
      \citet{chen12} to those presented here (and labeled ``near-IR").
      Every panel plots the difference in $\log M_*$ against near-IR $M_*$ in the left column, and against redshift in
      the right column. The 3$\sigma$-clipped averages (light grey circles) and 1$\sigma$ standard deviation (gold
      lines) of the difference distributions are over-plotted.   Shaded contours with levels separated by 0.3 dex
      increases in data density are displayed.  The comparison is restricted to $\log M_*/\msun > 11.2$ for the
      right-hand panels.  Overall normalization differences at the 0.1 dex level are expected.  Of greater importance
      are possible systematic \emph{trends} that may confuse evolutionary interpretations.  Differences in $M_*$ estimates display little or no systematic trends with $M_*$, but biases,
      likely resulting from a number of factors, are more apparent as a function of redshift.  See text for discussion.
      \label{fig:masscomp_portpca}}

    \end{figure*}

    \begin{figure*}
    \epsscale{1.2}
    \plotone{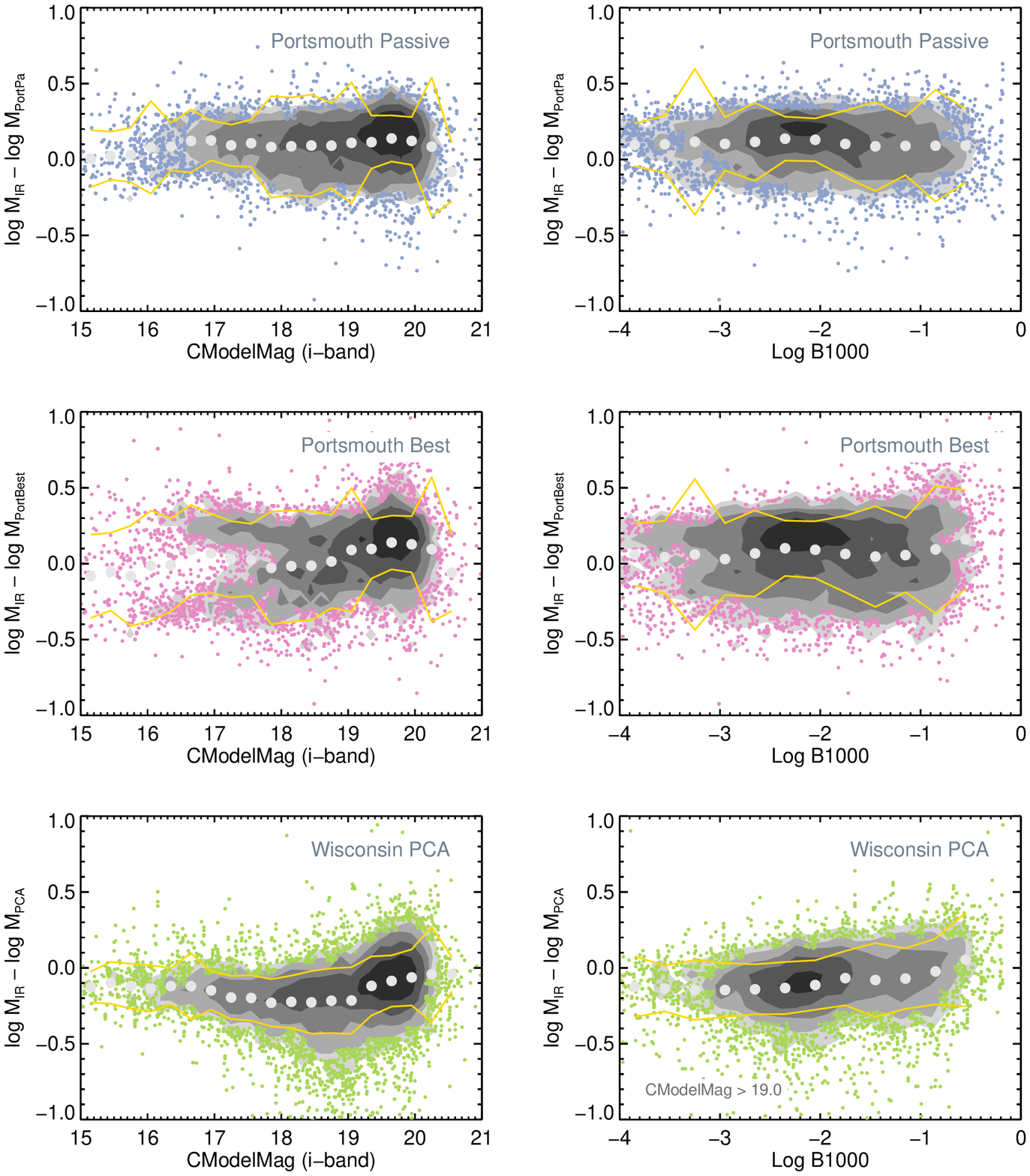}

    \caption{Mass comparisons as in Figure \ref{fig:masscomp_portpca} plotted now as a function of observed magnitude and the \birth{}
    parameter, a proxy for the degree of recent star formation less sensitive to dust than an NUV-optical color.  Some
    of the redshift-dependent trends in the $M_*$ offsets evident in Figure \ref{fig:masscomp_portpca} may result from
    a more fundamental dependence on magnitude.  The comparisons reveal little dependence on \birth{}, despite the fact
    that the near-IR \cat{} $M_*$ values are less sensitive to biases in $M/L$ owing to recent star formation.  The
    lack of any clear dependence on \birth{} in the Portsmouth Passive comparison is particularly striking.  Galaxies
    with higher \birth{} values tend to be at greater redshifts, thus explaining the \birth{}-dependent trend in the
    Wisconsin comparison.
    \label{fig:masscomp_mag}}

    \end{figure*}

\subsubsection{Extant BOSS $M_*$ estimates}

Systematic uncertainties are commonly acknowledged in studies employing $M_*$ estimates but not always carefully
studied.  The largest uncertainty is the IMF assumption which, to first order, can cause factor of $\sim$2 offsets in
$M_*$ values.  Setting the IMF aside, \citet{moustakas13} provide a careful examination at how different stellar
population models and assumptions regarding priors can influence $M_*$ estimates based on the same set of photometry.
While
\citet{moustakas13} demonstrate that the conclusions in their paper are largely insensitive to these systematic
effects, we adopt a similar methodology in Paper III that reveals the much greater importance of adopted priors and
assumptions for constraining massive galaxy evolution with high precision.

This section presents a preview of Paper III by comparing the fiducial near-IR $M_*$ estimates described above to
publicly available\footnote{\texttt{https://www.sdss3.org/dr10/spectro/galaxy.php}} estimates for BOSS galaxies from
the SDSS-III collaboration.  For a more systematic study of why $M_*$ offsets can occur between estimators (including
differences in the basis photometry set, stellar synthesis models, and adopted priors), please see Paper III.

In the current work, we only make comparisons to previously (spectroscopically) measured BOSS and SDSS Legacy galaxies
that use either the single-epoch SDSS-only photometry or SDSS spectroscopy.  We compare this dataset to the ``Wisconsin PCA'' masses
\citep{chen12}, two versions of the ``Portsmouth'' masses described in \citet[][referred to hereafter as
M13]{maraston13}, and the ``Granada'' masses released in SDSS DR10 \citep{ahn14}.  The Granada and Portsmouth masses
are based on stellar population synthesis SED fitting to the single-epoch SDSS $ugriz$ \texttt{ModelMag} colors (after
correcting for galactic extinction).  The Wisconsin PCA masses are derived from a PCA analysis of optical wavelength
regions of the BOSS spectra.  The resulting $M_*/L$ ratios in all three cases are scaled to total magnitudes as given
by the
\texttt{CModelMag\_i} measurements. All BOSS estimates  assume a flat cosmology with $\Omega_m = 0.274$ and $H_0 = 70$
km s$^{-1}$ Mpc$^{-1}$.  The impact on $M_*$ from using these cosmological parameters compared to
those adopted for the \cat{} $M_*$ estimates is less than 0.01 dex.

          \begin{figure*}
          \epsscale{1.2}
          \plotone{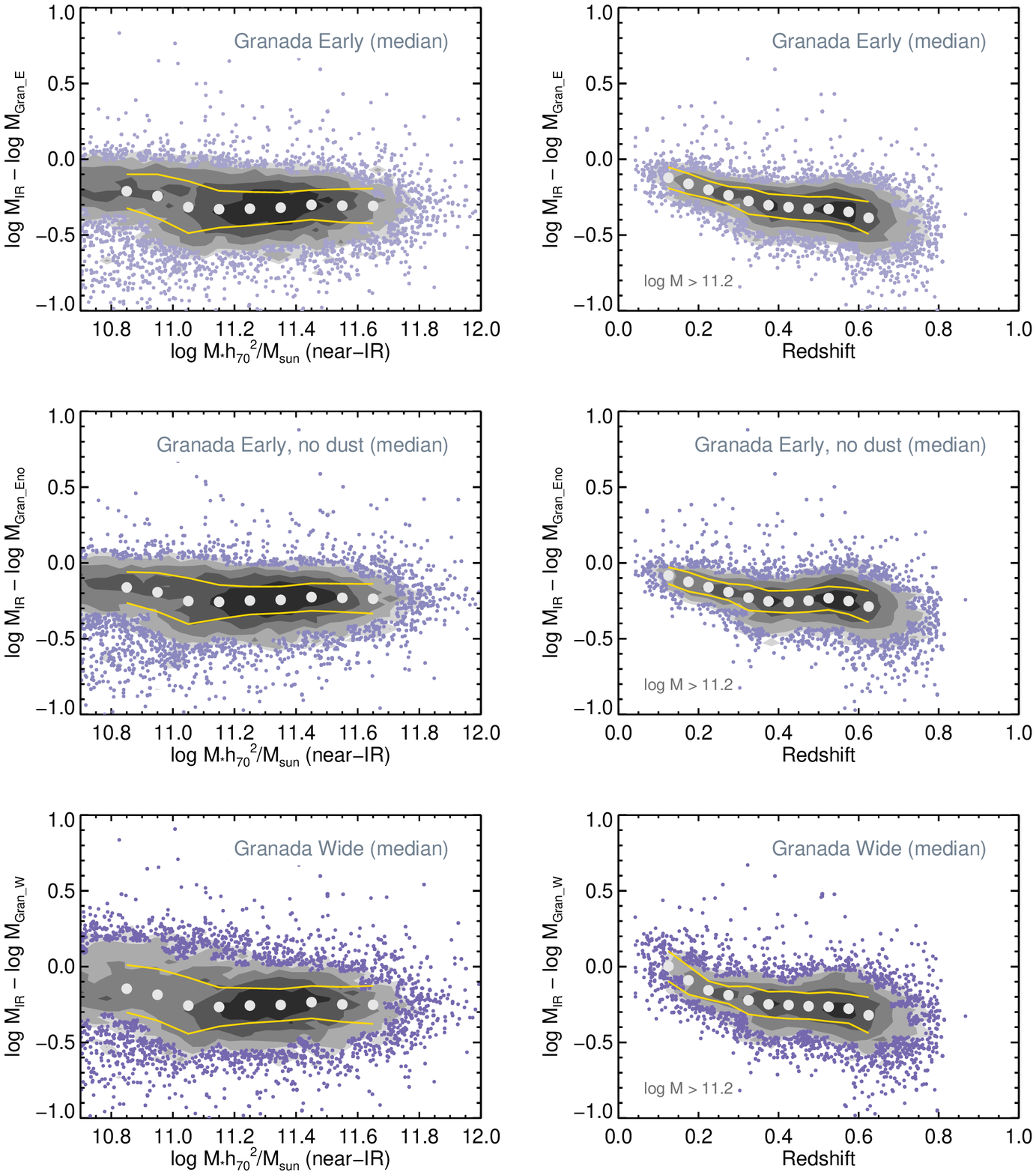}

    \caption{Mass comparisons, as in Figure \ref{fig:masscomp_portpca}, to the Granada $M_*$ estimates.  We consider
    the median value of the Granada $M_*$ posterior distribution.  The comparisons show relatively tight scatter but
    strong redshift-dependent trends that may be induced by a discrepancy in luminosity normalization, as evident in
    the more uniform trend with observed magnitude presented in Figure \ref{fig:masscomp_granmag}. \label{fig:masscomp_gran}}
          \end{figure*}

          \begin{figure*}
          \epsscale{1.2}
          \plotone{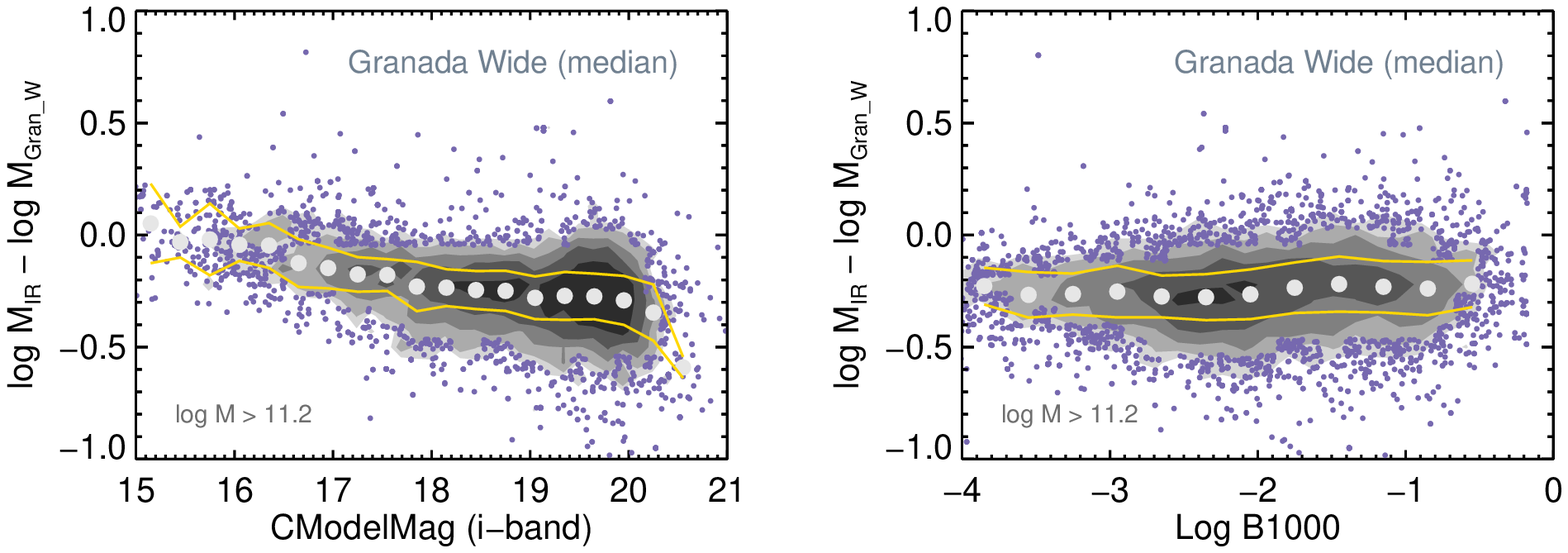}

    \caption{Granada $M_*$ comparisons as a function of observed magnitude and the \birth{} parameter for the Granada
    Wide estimates, which are the most similar in adopted priors to the \cat{} measurements.  The relative
    increase in the Granada $M_*$ at faint magnitudes continues smoothly across the full magnitude range, suggesting a
    difference in luminosity normalization with the \cat{} estimates. \label{fig:masscomp_granmag}}
          \end{figure*}

The ``passive'' only version of the Portsmouth masses are derived by finding the best fit between the $ugriz$
photometry and a single SED template built from the combination of two passively evolving, coeval bursts of star
formation.  The first burst is given solar metallicity and accounts for 97{\%} of the total mass, while the second,
accounting for the remainder, has 0.05 solar metallicity.  Age is the only fitting parameter for this ``passive"
template but ages less than 3 Gyr are not allowed.  A second set of M13
estimates, referred to as ``SF", are based upon fitting stellar population models with a range of star formation histories,
ages, and metallicities.  Both exponentially declining and truncated star formation histories are included, and an
upward correction of +0.25 dex is added based on an analysis of simulated data.  Neither of the M13 estimates includes
possible dust extinction inside the target galaxy. We use DR10 versions\footnote{\tt
https://www.sdss3.org/dr10/spectro/galaxy\_portsmouth.php} with Kroupa IMFs for the M13 estimates.  As with the \cat{}
estimates described in Section \ref{KBmass}, both M13 estimates measure the {\em current} mass in stars and stellar
remnants at the age of the stellar population model.
\citet{maraston13} combine the passive and SF estimates into a single, ``best" estimate by choosing the passive
estimate for galaxies with $(g-i) > 2.35$ and the SF estimate for $(g-i) < 2.35$.  In comparisons described below, we
 attempt to reproduce the M13 ``best'' estimates by adopting the passive or SF $M_*$ value from the Portsmouth catalogs
 using the Coadd $(g-i)$ color.  There will be slight differences from the actual $M_*$ values used in M13 because the
 Coadd photometry is deeper.

For the Wisconsin PCA masses, \citet{chen12} compared their PCA eigenspectra to BC03 models and assume a Kroupa IMF. We
only make comparisons for sources with PCA ``warning'' values set to zero.  The PCA mass estimates suffer
from aperture biases, which are increasingly important at lower redshifts, and from biases induced by low S/N CMASS
spectra which become significant at higher redshifts.  Unique to any of the estimates compared here, the Wisconsin
masses include priors on the presence of stochastic burst episodes (parameterized by the burst amplitude or mass
fraction and a burst duration time) of constant star formation and the possibility of random, abrupt truncations in the
star formation history.

The Granada estimates are based on SED fits to the FSPS models \citep[FSPS: Flexible Stellar Population
Synthesis][]{conroy09,conroy10a} and a Bayesian approach similar to that adopted for our fiducial estimates.  Four sets
of priors are adopted which change the nature of ranges allowed for metallicity, star formation history (parameterized
with $\tau$ models), age (less than the cosmic age), and dust content.  The ``Early'' priors restrict to
non-star-forming templates. ``Wide'' priors allow models with more recent star formation.  Both of these are available
with and without potential dust extinction.  We compare our results to Early (with dust), Early (no dust) which is
similar to the M13 template, and Wide with dust (which is most similar to our fiducial priors).  There are no burst
models in the Granada templates.  Both median and peak values of the $M_*$ posterior are provided.  We use the median
reported values, but choosing ``mean'' values does not make a significant difference, although the scatter increases
for the ``wide'' set of priors.

Comparisons of the Portsmouth and PCA masses to our near-IR \cat{} estimates are shown both as a function
of near-IR $M_*$ and redshift in Figure \ref {fig:masscomp_portpca}.  Figure \ref{fig:masscomp_mag} presents the
same comparisons against observed magnitude and the \birth{} birth parameter (Section \ref{birth}).  A similar set of
comparisons against the Granada masses is displayed in Figures \ref{fig:masscomp_gran} and \ref{fig:masscomp_granmag}.

\subsubsection{$M_*$ comparison analysis}

Despite different prior assumptions compared to the \cat{} $M_*$ estimates and the use of different observational data
sets, Figures \ref{fig:masscomp_portpca} through \ref{fig:masscomp_granmag} demonstrate agreement between the \cat{},
Portsmouth, Wisconsin, and Granada estimates at the 0.1 dex level, with scatter in the comparisons of 0.2--0.3 dex that
is consistent with the expected 0.1--0.2 dex uncertainties associated with any one estimator.  As we explore possible
trends below, it is important to emphasize that no single estimator can be considered ``truth.''  The \cat{} estimates,
however, benefit from more precise SEDs (thanks to the deeper Coadd photometry) and, because they are based on the
near-IR, are less subject to potential biases from recent star formation.  We emphasize that the comparisons discussed
here are restricted to bright, BOSS galaxies only.  An important advantage of the \cat{} is the ability to characterize
galaxies fainter than BOSS and outside the BOSS selection boundaries in order to build $M_*$-complete samples.  

In general, overall systematic offsets between $M_*$ estimates are tolerable (and expected), but as shown in Paper III,
the precision afforded by data sets such as \cat{} now require understanding subtle dependencies in offsets at the 0.1
dex level.  As a function of redshift or galaxy type, systematic biases or changes in the scatter at this level can
strongly impact conclusions on evolutionary trends.

Focusing first on the Portsmouth Passive masses, we see a relatively tight relation (upper panels in both figures)
against the \cat{} estimates with a hint of increasing scatter at high \birth{} values as might be expected if the
passive template breaks down for galaxies with more recent star formation.  The Portsmouth Best estimates feature more
scatter and structure, perhaps revealing how the combination of star-forming templates with a passive
template produces a degree of disjoint structure in some regions of parameter space.

The Wisconsin PCA masses feature a tight primary relation with the \cat{} estimates plus a secondary distribution that
scatters downward in the lower panels of Figure \ref{fig:masscomp_portpca}.  Figure
\ref{fig:masscomp_mag} (bottom-left panel) suggests this scatter towards larger PCA $M_*$ values derives from $i \gtrsim 19$ sources where the BOSS
spectral S/N drops and \citet{chen12} report biases in their own comparisons to SED-based $M_*$ estimates (although the
sense of the bias appears to be reversed).  A similar bias in the reverse direction was reported in M13.  A gentle
dependence on \birth{} (roughly 0.1 dex across the sample) in the Wisconsin comparison is also visible, highlighting
the effect of different priors in treating galaxies with more recent star formation (and higher dust extinction).
Because the fraction of galaxies with higher \birth{} values increases with redshift, this trend may underly the
redshift dependence in Wisconsin $M_*$ offset at $z > 0.4$ (Figure
\ref{fig:masscomp_portpca}, bottom-right panel).  As the redshift decreases below $z
\sim 0.4$, we speculate that aperture effects may drive the Wisconsin masses to lower values compared to the \cat{}
estimates. 

Turning now to the Granada masses, Figures \ref{fig:masscomp_gran} and \ref{fig:masscomp_granmag} show some of the
tightest relations (especially when plotted as a function of redshift, right-hand panels in Figure
\ref{fig:masscomp_gran}) compared to the \cat{} estimates.  This behavior may not be surprising given the similarity in
priors used by Granada compared to the \cat{}.  The level of scatter is comparable to the $z \lesssim 0.4$ regime of
the Wisconsin PCA comparison, which is encouraging because the Wisconsin group included bursts and truncated SFHs, but
otherwise similar ranges for more continuous SFHs and stellar population parameters.

Despite the tighter scatter, there are strong redshift-dependent trends in the Granada
comparisons with $z\sim 0.6$ galaxies about 0.3 dex more massive compared to the \cat{} estimates than those at $z\sim
0$.  This trend flattens somewhat at $z \gtrsim 0.4$, but appears to continue when plotted as a function of observed
magnitude (Figure \ref{fig:masscomp_granmag}, left panel) suggesting a discrepancy in the overall normalization of the
luminosity.  Understanding this discrepancy will be the subject of future work, but the lack of systematic trends as a
function of $M_*$ or \birth{} is nonetheless encouraging.

\section{Summary and Conclusions}\label{summary}

In this first paper in a series studying massive galaxies in Stripe 82, we have presented a new, publicly available
compilation of Stripe 82 data sets, the Stripe 82 Massive Galaxy Catalog (\cat{}), designed to enable $M_*$-limited
studies of massive galaxy evolution since $z \sim 0.7$.  The catalog includes 9-band $ugrizYJHK$ photometry obtained by
matching the SDSS Coadd and UKIDSS-LAS photometric catalogs using the \synmag{} package.  Exploiting over 149,439
spectroscopic redshifts, we have assembled and tested a set of photometric redshifts from a variety of sources that are
needed to complement \speczs{} in order to build $M_*$ complete samples.  With this redshift information, we have
derived a new set of near-IR based $M_*$ estimates and included them in the \cat{}.

A number of additional steps have been taken to make the \cat{} ready for population studies.  We have used the near-IR
photometry to improve the star-galaxy separation and built new estimates of UKIDSS total magnitudes that address the
significant limitations due to blends in the public UKIDSS-LAS catalogs.  We have also carefully defined the survey
geometry and local depth of the combined data set, designing new custom masks in order to reject problematic regions in
both the SDSS Coadd and UKIDSS imaging bands.

Applying these rejection masks and a set of near-IR magnitude limits, we have constructed a $M_*$-limited sub-sample of
the \cat{} called \ukwide{} that spans \area{} deg$^2$ and contains 41,770 galaxies with $\logm > 11.2$ to $z \approx
0.7$, roughly 45\% of which boasts a spectroscopic redshift.  This is the largest near-IR selected and $M_*$ complete
sample of galaxies beyond $z \sim 0.1$ assembled to date.  Paper II exploits this sample to study the completeness
function in the BOSS samples.

Finally, we have presented comparisons between the near-IR based $M_*$ estimates in the \cat{} and previous, publicly
available $M_*$ estimates for BOSS galaxies.  We analyze various systematic trends that are apparent, and find
generally good agreement at the 0.1 dex level.  However, as demonstrated in our forthcoming analysis of galaxy $M_*$
functions (Paper III), systematics at this level are now a significant limitation in our ability to measure high
precision evolution in the massive galaxy population. Encouragingly for the BOSS sample, the addition of near-IR data
has only a mild impact on $M_*$ estimates, primarily affecting galaxies with more recent star formation.

\section{Acknowledgments}

This work was supported by World Premier International Research Center Initiative (WPI Initiative), MEXT, Japan.  This
work was supported by a Kakenhi Grant-in-Aid for Scientific Research 24740119 from Japan society for the Promotion of
Science.  We thank E.~Rykoff and E.~Rozo for a generous contribution of \redmapper{} photometric redshift estimates.
This publication has made use of code written by James R. A. Davenport.  We are grateful to Aur\'elien Benoit-L\'evy
for help with setting \mangle{} parameters.

Funding for SDSS-III has been provided by the Alfred P. Sloan Foundation, the Participating Institutions, the National Science Foundation, and the U.S. Department of Energy Office of Science. The SDSS-III web site is \texttt{http://www.sdss3.org/}.

SDSS-III is managed by the Astrophysical Research Consortium for the Participating Institutions of the SDSS-III Collaboration including the University of Arizona, the Brazilian Participation Group, Brookhaven National Laboratory, Carnegie Mellon University, University of Florida, the French Participation Group, the German Participation Group, Harvard University, the Instituto de Astrofisica de Canarias, the Michigan State/Notre Dame/JINA Participation Group, Johns Hopkins University, Lawrence Berkeley National Laboratory, Max Planck Institute for Astrophysics, Max Planck Institute for Extraterrestrial Physics, New Mexico State University, New York University, Ohio State University, Pennsylvania State University, University of Portsmouth, Princeton University, the Spanish Participation Group, University of Tokyo, University of Utah, Vanderbilt University, University of Virginia, University of Washington, and Yale University.

\clearpage
\appendix
\section{UKIDSS Query}

\lstset{language=SQL, basicstyle=\footnotesize}
\begin{lstlisting}
SELECT s.sourceID, s.RA, s.dec, s.mergedClassStat, s.mergedClass, 
s.pStar, s.pGalaxy, s.pNoise, s.pSaturated, s.eBV, s.aY, s.aJ, s.aH, s.aK,

s.yPetroMag, s.yPetroMagErr, s.yHallMag, s.yHallMagErr, ydtcn.IsoMag AS YIsoMag, s.yppErrBits,
s.yAperMag3, s.yAperMag3Err, s.yAperMag4, s.yAperMag4Err, s.yAperMag6, s.yAperMag6Err, 
ymd.aperCor3 AS yaperCor3, ymd.aperCor4 AS yaperCor4, ymd.aperCor6 AS yaperCor6,

ymd.seeing AS yseeing, ymd.skyCorrCat as ySkyCorrCat, 
ymd.photZPCat as yPhotZPCat, ymd.photZPErrCat as yPhotZPErrCat,
ydtcn.aperMag1 AS yaperMag1, ydtcn.AperMag1Err AS yAperMag1Err, ymd.aperCor1 as yaperCor1, 
ydtcn.aperMag2 AS yaperMag2, ydtcn.AperMag2Err AS yAperMag2Err, ymd.aperCor2 as yaperCor2, 
ydtcn.aperMag5 AS yaperMag5, ydtcn.AperMag5Err AS yAperMag5Err, ymd.aperCor5 as yaperCor5, 
ydtcn.ell AS yEll, ydtcn.pHeight AS ypHeight, ydtcn.pHeightErr AS ypHeightErr, 

s.j_1PetroMag, s.j_1PetroMagErr, s.j_1HallMag, s.j_1HallMagErr, j_1dtcn.IsoMag AS J_1IsoMag, 
s.j_1ppErrBits,
s.j_1AperMag3, s.j_1AperMag3Err, s.j_1AperMag4, s.j_1AperMag4Err, s.j_1AperMag6, s.j_1AperMag6Err, 
j_1md.aperCor3 AS j_1aperCor3, j_1md.aperCor4 AS j_1aperCor4, j_1md.aperCor6 AS j_1aperCor6, 

j_1md.seeing AS j_1seeing, j_1md.skyCorrCat as j_1SkyCorrCat, 
j_1md.photZPCat as j_1PhotZPCat, j_1md.photZPErrCat as j_1PhotZPErrCat,
j_1dtcn.aperMag1 AS j_1aperMag1, j_1dtcn.AperMag1Err AS j_1AperMag1Err, j_1md.aperCor1 as j_1aperCor1, 
j_1dtcn.aperMag2 AS j_1aperMag2, j_1dtcn.AperMag2Err AS j_1AperMag2Err, j_1md.aperCor2 as j_1aperCor2, 
j_1dtcn.aperMag5 AS j_1aperMag5, j_1dtcn.AperMag5Err AS j_1AperMag5Err, j_1md.aperCor5 as j_1aperCor5, 
j_1dtcn.ell AS j_1Ell, j_1dtcn.pHeight AS j_1pHeight, j_1dtcn.pHeightErr AS j_1pHeightErr, 

s.hPetroMag, s.hPetroMagErr, s.hHallMag, s.hHallMagErr, hdtcn.IsoMag AS HIsoMag, s.hppErrBits,
s.hAperMag3, s.hAperMag3Err, s.hAperMag4, s.hAperMag4Err, s.hAperMag6, s.hAperMag6Err, 
hmd.aperCor3 AS haperCor3, hmd.aperCor4 AS haperCor4, hmd.aperCor6 AS haperCor6,

hmd.seeing AS hseeing, hmd.skyCorrCat as hSkyCorrCat, 
hmd.photZPCat as hPhotZPCat, hmd.photZPErrCat as hPhotZPErrCat,
hdtcn.aperMag1 AS haperMag1, hdtcn.AperMag1Err AS hAperMag1Err, hmd.aperCor1 as haperCor1, 
hdtcn.aperMag2 AS haperMag2, hdtcn.AperMag2Err AS hAperMag2Err, hmd.aperCor2 as haperCor2, 
hdtcn.aperMag5 AS haperMag5, hdtcn.AperMag5Err AS hAperMag5Err, hmd.aperCor5 as haperCor5, 
hdtcn.ell AS hEll, hdtcn.pHeight AS hpHeight, hdtcn.pHeightErr AS hpHeightErr, 

s.kPetroMag, s.kPetroMagErr, s.kHallMag, s.kHallMagErr, kdtcn.IsoMag AS KIsoMag, s.kppErrBits,
s.kAperMag3, s.kAperMag3Err, s.kAperMag4, s.kAperMag4Err, s.kAperMag6, s.kAperMag6Err, 
kmd.aperCor3 AS kaperCor3, kmd.aperCor4 AS kaperCor4, kmd.aperCor6 AS kaperCor6,

kmd.seeing AS kseeing, kmd.skyCorrCat as kSkyCorrCat, 
kmd.photZPCat as kPhotZPCat, kmd.photZPErrCat as kPhotZPErrCat,
kdtcn.aperMag1 AS kaperMag1, kdtcn.AperMag1Err AS kAperMag1Err, kmd.aperCor1 as kaperCor1, 
kdtcn.aperMag2 AS kaperMag2, kdtcn.AperMag2Err AS kAperMag2Err, kmd.aperCor2 as kaperCor2, 
kdtcn.aperMag5 AS kaperMag5, kdtcn.AperMag5Err AS kAperMag5Err, kmd.aperCor5 as kaperCor5, 
kdtcn.ell AS kEll, kdtcn.pHeight AS kpHeight, kdtcn.pHeightErr AS kpHeightErr, 

l.ymfID as ymfID, l.yeNum as yeNum, l.j_1mfID as j_1mfID, l.j_1eNum as j_1eNum, l.j_2mfID as j_2mfID, 
l.j_2eNum as j_2eNum, l.hmfID as hmfID, l.heNum as heNum, l.kmfID as kmfID, l.keNum as keNum

FROM 
  lasSource AS s, lasMergeLog AS l, 
  MultiframeDetector AS ymd, MultiframeDetector AS j_1md, MultiframeDetector AS j_2md, 
  MultiframeDetector AS hmd, MultiframeDetector AS kmd,
  lasDetection AS ydtcn, lasDetection AS j_1dtcn, lasDetection AS j_2dtcn, 
  lasDetection AS hdtcn, lasDetection AS kdtcn
WHERE
  s.frameSetID = l.frameSetID AND
  l.ymfID = ymd.multiframeID AND
  l.yeNum = ymd.extNum AND
  l.j_1mfID = j_1md.multiframeID AND
  l.j_1eNum = j_1md.extNum AND
  l.j_2mfID = j_2md.multiframeID AND
  l.j_2eNum = j_2md.extNum AND
  l.hmfID = hmd.multiframeID AND
  l.heNum = hmd.extNum AND
  l.kmfID = kmd.multiframeID AND
  l.keNum = kmd.extNum AND

  l.ymfID = ydtcn.multiframeID AND
  l.yeNum = ydtcn.extNum AND
  s.ySeqNum = ydtcn.SeqNum AND

  l.j_1mfID = j_1dtcn.multiframeID AND
  l.j_1eNum = j_1dtcn.extNum AND
  s.j_1SeqNum = j_1dtcn.SeqNum AND

  l.j_2mfID = j_2dtcn.multiframeID AND
  l.j_2eNum = j_2dtcn.extNum AND
  s.j_2SeqNum = j_2dtcn.SeqNum AND

  l.hmfID = hdtcn.multiframeID AND
  l.heNum = hdtcn.extNum AND
  s.hSeqNum = hdtcn.SeqNum AND

  l.kmfID = kdtcn.multiframeID AND
  l.keNum = kdtcn.extNum AND
  s.kSeqNum = kdtcn.SeqNum AND
  (s.priOrSec <= 0 OR s.priOrSec = s.frameSetID) AND
  s.dec > -1.3 AND s.dec < 1.3 AND (s.ra between 0 AND 30)


\end{lstlisting}

\bigskip
\section{\cat{} Data Products}

Several types of \cat{} data products are made publicly available at {\texttt
{s82mgc.massivegalaxies.com}}.  Documentation on the website provides the most up-to-date
description of these products.  We summarize what is available below.  In some cases to make file sizes more
manageable, we have divided Stripe 82 into east and west sections by splitting at $\alpha_{\rm
J2000}=0$.  

\medskip
\noindent \textbf{Parent Catalogs}

Parent photometric catalogs in FITS table format for the east and west side of Stripe 82 are provided from queries of
the SDSS Coadd and UKIDSS LAS databases.  The Coadd parent catalogs have filenames beginning with ``\texttt{S82coadd\_}''
and contain the Coadd photometry and SDSS profile fitting results.  The UKIDSS catalogs have filenames beginning with
``\texttt{las\_DR8\_}'' and contain the UKIDSS photometry, error flags, seeing, aperture corrections, and other information
from the UKIDSS LAS database.

\medskip
\noindent \textbf{The Stripe 82 Massive Galaxy Catalog}

The heart of the \cat{} itself, with a filename starting with ``\texttt{pcatd\_}'', is selected on the SDSS Coadd
parent catalog to include Coadd sources with measured $r$-band profile fit information (in practice sources with
\texttt{devAB\_R} and
\texttt{ExpAB\_R} $>$ 0.01 are selected).  The \texttt{pcat} table includes cross-referencing information for the UKIDSS catalog,
matched SDSS+UKIDSS \synmag{} photometry, and corrected UKIDSS total magnitudes, among other selected information from both
parent catalogs.  The \texttt{pcat} FITS table is a single file that contains 15,342,585 with 92 tags for a total size
of 5.5 GB.

\medskip
\noindent \textbf{Galaxy Properties}

Additional FITS tables, matched one-to-one to the \texttt{pcat}, provide estimates of galaxy properties for sources in
the \cat{}.  These include spectroscopic redshifts from SDSS, VVDS, and DEEP2 and all photometric redshifts discussed
in Section \ref{photoz}.  The \cat{} fiducial $M_*$ estimates as well as \kcorrect{} absolute magnitudes and \birth{}
values are also provided.  Please see the website for details.

\medskip
\noindent \textbf{The \ukwide{} sample}

For those interested in working with the $M_*$-limited \ukwide{} galaxy sample, smaller sub-catalogs of the
\texttt{pcat} and galaxy property tables are provided.  Some software tools needed to construct \ukwide{}, e.g., the
star-galaxy separation, are also included.

\medskip
\noindent \textbf{Survey Footprint}

The UKIDSS depth information and all \mangle{} polygon files describing the geometric layout of the different survey
components as well as the rejection masks are provided and described on the website.

\bibliographystyle{apj}
\bibliography{/Users/kbundy/Documents/bibliographies/references.bib}

\end{document}